\RequirePackage{fix-cm}
\documentclass{svjour3}                     % onecolumn (standard format)
\smartqed  % flush right qed marks, e.g. at end of proof
\usepackage{graphicx}
\usepackage[utf8x]{inputenc}
\usepackage[english]{babel}
\usepackage{natbib}
\usepackage{amsmath}
\usepackage{amsfonts}

\newcommand{\T}{^\mathrm T}
\newcommand{\argmax}{\mathrm{argmax}}
\begin{document}

\title{Weighted likelihood mixture modeling and model based clustering}


%\titlerunning{WLE mixture modelling}        % if too long for running head

\author{Luca Greco   \and Claudio Agostinelli      
         %etc.
}


\institute{L. Greco \at
              DEMM Department\\ University of Sannio, Italy \\
              {\tt luca.greco@unisannio.it}
                      %  \\
           \and
           C. Agostinelli \at
              Department of Mathematics\\ University of Trento, Italy \\
              \email{claudio.agostinelli@unitn.it}  
}

\date{Received: date / Accepted: date}

\maketitle

\begin{abstract}
A weighted likelihood approach for robust fitting of a mixture of multivariate Gaussian components is developed in this work. 
Two approaches have been proposed that are driven by a suitable modification of the standard EM and CEM algorithms, respectively. In both techniques, the M-step is enhanced by the computation of weights aimed at downweighting outliers. The weights are based on Pearson residuals stemming from robust Mahalanobis-type distances.
Formal rules for robust clustering and outlier detection can be also defined based on the fitted mixture model.  
The behavior of the proposed methodologies has been investigated by some numerical studies and real data examples in terms of both fitting and classification accuracy and outlier detection.
\end{abstract}

\section{Introduction}
Multivariate normal mixture models represent a very popular tool for both density estimation and clustering \citep{mclachlan2004finite}. 
The parameters of a mixture model are commonly estimated by maximum likelihood by resorting to the EM algorithm \citep{dempster1977maximum}. Let $y=(y_1,y_2,\ldots,y_n)^{\T}$ be a random sample of size $n$.
The mixture likelihood can be expressed as
\begin{equation}
\label{mixlik}
L(y; \tau)=\prod_{i=1}^n\sum_{k=1}^K \pi_k\phi_p(y_i; \mu_k,\Sigma_k) \ ,
\end{equation}
where $\tau=(\pi, \mu_1, \ldots,  \mu_K,  \Sigma_1, \ldots, \Sigma_K)$, $\phi_p(\cdot;\cdot)$ is the $p$-dimensional multivariate normal density, $\pi=(\pi_1,\ldots,\pi_K)$ denotes the vector of prior membership probabilities and $(\mu_k,\Sigma_k)$ are the mean vector and variance-covariance matrix of the $k^{th}$ component, respectively. 
Rather than using the likelihood in (\ref{mixlik}), the EM algorithm works with the complete likelihood function 
\begin{equation}
\label{complik}
L^c(y;\tau)=\prod_{i=1}^n\prod_{k=1}^K \left[\pi_k\phi_p(y_i; \mu_k,\Sigma_k) \right]^{u_{ik}}\ ,
\end{equation}
where $u_{ik}$ is an indicator of the $i^{th}$ unit belonging to the $k^{th}$ component. 
The EM algorithm iteratively alternates between two steps: expectation (E) and maximization (M).
In the E-step, the posterior expectation of (\ref{complik}) is evaluated by setting $u_{ik}$ equal to the posterior probability that $y_i$ belongs to the $k^{th}$ component, i.e.
$$
u_{ik}\propto \pi_k\phi_p(y_i; \mu_k,\Sigma_k) \ , 
$$
whereas at the M step $\pi$, $\mu_k$ and $\Sigma_k$ are estimated conditionally on $u_{ik}$.

An alternative strategy is given by the penalized Classification EM (CEM) algorithm \citep{symon1977clustering, bryant1991large, celeux1993comparison}: the substantial difference is that the E-step is followed by a C-step (where C stands for classification) in which $u_{ik}$ is estimated as either $0$ or $1$, meaning that each unit is assigned to the most likely component, conditionally on the current parameters' values, i.e.
$k_i=\argmax_k u_{ik}$, $u_{ik_i}=1$ and $u_{ik}=0$ for $k\neq k_i$.
%Each iteration of the CEM is made of one E-step, C-step and M-step.
The classification approach is aimed at maximizing the corresponding classification likelihood (\ref{complik}) over both the mixture parameters and the individual components' labels. 
In the case $\pi_k=1/K$, then the standard CEM algorithm is recovered. A detailed comparison of the EM and CEM algorithms can be found in \cite{celeux1993comparison}.

When the sample data are prone to contamination and several unexpected outliers occur with respect to (w.r.t.) the assumed mixture model, maximum likelihood is likely to lead to unrealistic estimates and to fail in recovering the underlying clustering structure of the data (see \cite{farcomeni2016robust} for a recent account).
In the presence of noisy data that depart from the underlying mixture model, there is the need to replace maximum likelihood with a suitable robust procedure, leading to estimates and clustering rules that are not badly affected by contamination.% and allow to recover the underlying clustering structure of the data at hand.

The need for robust tools in the estimation of mixture models has been first addressed in \cite{campbell1984mixture}, who suggested to replace standard maximum likelihood with M-estimation.
In a more general fashion, \cite{farcomeni2015s} proposed to resort to multivariate S-estimation of location and scatter in the M-step. Actually, the authors focused their attention on Hidden Markov Models but their approach can be adapted from dynamic to static finite mixtures. According to such strategies, each data point is attached a weight lying in $[0,1]$ (a strategy commonly addressed as soft trimming). An alternative approach to robust fitting and clustering is based on hard trimming procedures, i.e. a crispy  weight $\left\{0, 1\right\}$ is attached to each observation: atypical observations are expected to be trimmed and the model is fitted by using a subset of the original data.
 The {\tt tclust} methodology \citep{garcia2008general, fritz2013fast} is particularly appealing. A very recent extension has been discussed in \cite{dotto2016reweighting}, who proposed a reweighted trimming procedure named  {\tt rtclust}.
Both the {\tt tclust} and {\tt rtclust} are based on penalized CEM algorithms modified by a trimming strategy: at each iteration the most distant data points, with distances computed conditionally on the current cluster assignments, are discarded and a trimmed classification likelihood is maximized.
A related proposal has been presented in \cite{neykov2007robust} based on the so called trimmed likelihood methodology.  
Furthermore, it is worth to mention that mixture model estimation and clustering can be also implemented by using the adaptive hard trimming strategy characterizing the Forward Search \citep{atkinson2013exploring}. % 

There are also different proposals aimed at being robust that are not based on soft or hard trimming procedures. Some of them are characterized by the use of flexible components in the mixture. The idea is that of embedding the Gaussian mixture in a supermodel: \cite{mclachlan2004finite} introduced a mixture of Student’s t distributions, a mixture of skewed Student’s t distributions has been proposed in \cite{lin2010robust} and \cite{lee2014finite}, whereas \cite{fraley1998many, fraley2002model} considered an additional component modeled as a Poisson process to handle noisy data (the method is available from package {\tt mclust} in {\tt R} \citep{fraleymclust}.
A robust approach, named {\tt otrimle}, has been proposed recently by \cite{coretto2016robust}, who considered the addition of an improper uniform mixture component to accomodate outliers.

We propose a robust version of both the EM and the penalized CEM algorithms to fit a mixture of multivariate Gaussian components based on soft trimming, in which weights are evaluated according to the weighted likelihood methodology \citep{markatou1998}. 
A first attempt in this direction has been pursued by \cite{markatou2000mixture}.
Here, that approach has been developed further and made more general leading to a newly established technique, in which
weights are based on the recent results stated in \cite{agostinelli2017weighted}.
The methodology leads to a robust fit and is aimed at providing both cluster assignment of genuine data points and outlier detection rules. Data points flagged as anomalous are not meant to be classified into any of the clusters.
Furthermore, a relevant aspect of our proposal is represented by the introduction of constraints, not considered in \cite{markatou2000mixture}, aimed at avoiding local or spurious solutions \citep{fritz2013fast}.

Some necessary preliminaries on weighted likelihood estimation are given in Section~\ref{back}. The weighted EM and penalized CEM algorithms are introduced in Section \ref{WEM} and classification and outlier detection rules are outlined. 
Some illustrative example based on simulated data are given in Section~\ref{ill}, whereas Section~\ref{mod} is devoted to model selection.  Numerical studies are presented in Section~\ref{num} and  real data examples are discussed in Section~\ref{real}.

\section{Background}
%Weighted likelihood mixture modeling
\label{back}
Let us assume a mixture model composed by $K$ heterogeneous multivariate Gaussian components, where $K$ is fixed in advance, with density function denoted by $m(y;\tau)=\sum_{j=1}^K\pi_j\phi_p(y_i;\mu_j, \Sigma_j)$. 
\cite{markatou2000mixture} suggested to work with the following weighted likelihood estimating equation (WLEE) in the M-step of the EM algorithm:
\begin{equation}
\sum_{i=1}^n w_i \sum_{j=1}^k u_{ij}\frac{\partial}{\partial \tau}\left[ \log\pi_j +\log\phi_p(y_i;\mu_j, \Sigma_j)\right]=0 \ .
\label{wlee_mark}
\end{equation}
 
We notice that maximum likelihood estimates are replaced by weighted estimates. The weights are defined as
\begin{equation}
\label{weight}
w_i=w(\delta(y_i)) = \frac{\left[A(\delta(y_i)) + 1\right]^+}{\delta(y_i) + 1} \ ,
\end{equation}
where $[\cdot]^+$ denotes the positive part, $\delta(y)$ is the Pearson residual function and 
 $A(\delta)$ is the Residual Adjustment Function (RAF, \cite{basu1994minimum}).
 The Pearson residual gives a measure of the agreement between the assumed model $m(y; \tau)$ and the data, that are summarized by a non-parametric density estimate $\hat m_n(y)=n^{-1}\sum_{i=1}^n k(y; y_i, h)$, based on a kernel $k(t; y, h)$ indexed by a bandwidth $h$, that is
 \begin{equation}
\label{residual}
\delta(y) = \frac{\hat m_n(y) }{m(y; \tau)}-1 \ ,
\end{equation}
with $\delta\in[-1, \infty)$.
In the construction of Pearson residuals, \cite{markatou2000mixture} suggested to use a smoothed model density in the continuous case, by using the same kernel involved in non-parametric density estimation (see \cite{basu1994minimum, markatou1998} for general results), i.e. $$m^*(y; \tau)=\int k(t;y,h)m(t;\tau)dt.$$ 

When the model is correctly specified, the Pearson residual function (\ref{residual}) evaluated at the true parameter value converges almost surely to zero, whereas, otherwise, for each value of the parameters, large Pearson residuals detect regions where the observation is unlikely to occur under the assumed model.
The weight function (\ref{weight}) can be choosen to be unimodal so that it declines smoothly as the residual $\delta(y)$ departs from zero. Hence, those observations lying in such regions are attached a weight that decreases with increasing Pearson residual.  Large Pearson residuals and small weights will correspond to data points that are likely to be outliers.
The RAF plays the role to bound the effect of large residuals on the fitting procedure, as well as the Huber and Tukey-bisquare function bound large distances in M-type estimation and we assume is such that $|A(\delta)|<|\delta|$.
Here, we consider the families of RAF based on the Power Divergence Measure
\begin{equation*}
A_{pdm}(\delta) = \left\{
\begin{array}{lc}
\tau \left( (\delta + 1)^{1/\tau} - 1 \right) & \tau < \infty \\
\log(\delta + 1) & \tau \rightarrow \infty 
\end{array}
\right .
\end{equation*}
Special cases are maximum likelihood ($\tau = 1$, as the weights become all equal to one), Hellinger distance ($\tau = 2$), Kullback--Leibler divergence ($\tau \rightarrow \infty$) and Neyman's Chi--Square ($\tau=-1$). 
 Another example is given by the Generalized Kullback-Leibler divergence (GKL) defined as $$A_{gkl}(\delta)=\log (\tau\delta+1)/\tau, \ 0\leq \tau \leq 1.$$ Maximum likelihood is a special case when $\tau\rightarrow 0$ and Kullback--Leibler divergence is obtained for $\tau=1$.

The shape of the kernel function has a very limited effect on weighted likelihood estimation. On the contrary, 
the smoothing parameter $h$ directly affects the robustness/efficiency trade-off of the methodology in finite samples. 
Actually, large values of $h$ lead to Pearson residuals all close to zero and weights all close to one and, hence, large efficiency, since the kernel density estimate is stochastically close to the postulated model. On the other hand, small values of $h$ make the kernel density estimate more sensitive to the occurrence of outliers and the Pearson residuals become large for those data points that are in disagreement with the model. In other words, in finite samples more smoothing will lead to higher efficiency but larger bias under contamination. 
%The smoothing parameter plays a minor role in the asymptotic properties of the procedure \citep{basu1994minimum}.

\section{Weighted likelihood mixture modeling}
\label{WEM}
The approach proposed by \cite{markatou2000mixture} exhibits the same limitations that have been highlighted in \cite{agostinelli2017weighted} in the case of weighted likelihood estimation of multivariate location and scatter. The main drawbacks are driven by the employ of multivariate kernels.
Actually, in the multivariate setting the computation of weights becomes troublesome because multivariate non-parametric density estimation is prone to the curse of dimensionality and the selection of $h$ becomes problematic, as well. 
The novel technique proposed in \cite{agostinelli2017weighted} is based on the 
Mahalanobis distances $$d=d(y;\mu,\Sigma)=[(y-\mu)^{\T}\Sigma^{-1}(y-\mu)]^{1/2}.$$ Then, Pearson residuals and weights are obtained by comparing a univariate kernel density estimate based on squared distances and their underlying (asymptotic) $\chi^2_p$ distribution at the assumed multivariate normal model, rather than working with multivariate data and multivariate kernel density estimates, i.e
 %In the following, we propose a weighted EM algorithm and penalized CEM algorithm characterized by 
%Pearson residuals defined as
\begin{equation}
\label{newP}
\delta(y)=\frac{\hat m_n\left(d^2\right)}{m_{\chi^2_p}\left(d^2\right)}-1 \ ,
\end{equation}
where 
\begin{equation*}
\hat m_n(t)=n^{-1}\sum_{i=1}^n k(t; d^2, h)
\label{kern}
\end{equation*}
 is an unbiased at the boundary univariate kernel density estimate over $(0, \infty)$ and $m_{\chi^2_p}(t)$ denotes the $\chi^2_p$ density function. 
It is worth noting that Pearson residuals can be evaluated w.r.t. the original $\chi^2_p$ density, so avoiding model smoothing (see also \cite{kuchibhotlabasu2015,kuchibhotlabasu2017a}). 

By exploiting the approach developed in \cite{agostinelli2017weighted}, we propose a weighted EM algorithm and penalized CEM algorithm whose M-steps are  
characterized by soft trimming procedures based on the Pearson residuals introduced in (\ref{newP}).

The Weighted EM algorithm (WEM) is structured as follows:
\begin{enumerate}
\item {\bf Initialization}: 
$$
\tau^{(0)}=(\pi^{(0)}, \mu_1^{(0)}, \ldots,  \mu_K^{(0)},  \Sigma_1^{(0)}, \ldots, \Sigma_1^{(K)}) \ .
$$
An appealing starting solution is given by a very robust method such as {\tt tclust} based on a large rate of trimming but other robust candidate solutions can be used to initialize the algorithm. In order to avoid the algorithm to be dependent on initial values, a simple and common strategy is to run the algorithm from a number (say, 20 to 50) of starting values.
For instance, different initial solutions can be obtained by randomly perturbing the deterministic starting solution and/or the final one obtained from it \citep{farcomeni2015s}. Details on how to choose the best solution will be given at the end of the Section.
%In this work we have obtained a total of 20 initial solutions in the real data application. For the numerical studies that follow, we have found that the de- terministic solution usually leads to the largest optimum for the likelihood.     
\item {\bf E step}: the standard E step is left unchanged, with 
$$
u_{ik}^{(s)}=\frac{\pi_k^{(s-1)}\phi_p\left(y_i; \mu_k^{(s-1)},\Sigma_k^{(s-1)}\right)}{\sum_{k=1}^K \pi_k^{(s-1)}\phi_p\left(y_i; \mu_k^{(s-1)},\Sigma_k^{(s-1)}\right)}
$$

\item {\bf Weighted M step}: based on current parameter estimates,
\begin{enumerate}
\item {\bf Soft-trimming}: let us evaluate component-wise Mahalanobis-type distances
$$
d_{ik}^{(s)}=d\left(y_i; \mu_k^{(s-1)}, \Sigma_k^{(s-1)}\right) \ .
$$
Then, for each group, compute Pearson residuals and weights as
$$
\delta_{ik}^{(s)}=\frac{\hat m_n\left(d_{ik}^{(s)^2}\right)}{m_{\chi^2_p}\left(d_{ik}^{(s)^2}\right)}-1
$$
and 
$$
w_{ik}^{(s)}=\frac{\left[A\left(\delta_{ik}^{(s)}\right)+1\right]^+}{\delta_{ik}^{(s)}+1}
$$
respectively. 

\item {\bf Update membership probabilities and component specific parameter estimates}:
\begin{eqnarray*}
\pi_k^{(s+1)}&=&\frac{\sum\limits_{i=1}^n u_{ik}^{(s)}w_{ik}^{(s)}}{\sum_{i=1}^n \sum_{k=1}^Ku_{ik}^{(s)}w_{ik}^{(s)}}\\
\mu_k^{(s+1)}&=&\frac{\sum\limits_{i=1}^n y_i w_{ik}^{(s)}u_{ik}^{(s)}}{\sum_{i=1}^n w_{ik}^{(s)}u_{ik}^{(s)}}\\
\Sigma_k^{(s+1)}&=&\frac{\sum\limits_{i=1}^n \left(y_i-\mu_k^{(s+1)}\right)\left(y_i-\mu_k^{(s+1)}\right)^{\T} w_{ik}^{(s)}u_{ik}^{(s)}}{\sum_{i=1}^n w_{ik}^{(s)}u_{ik}^{(s)}}
\end{eqnarray*} 
\item {\bf Set} $\tau^{(s+1)}=\left(\pi^{(s+1)}, \mu_1^{(s+1)}, \ldots,  \mu_K^{(s+1)},  \Sigma_1^{(s+1)}, \ldots, \Sigma_K^{(s+1)}\right)$ \ .
\end{enumerate}

\end{enumerate}
It is worth noting that at the M-step it is proposed to solve the following WLEE
\begin{equation}
\sum_{i=1}^n \sum_{j=1}^k u_{ij}\frac{\partial}{\partial \tau}\left[ \log\pi_j +\log\phi_p(y_i;\mu_j, \Sigma_j)\right]w_{ij} \ ,
\label{wlee_wem}
\end{equation}
that is characterized by the evaluation of $K$ component-wise sets of weights, rather than one weight for each observation, as in equation (\ref{wlee_mark}). 
 
The Weighted penalized CEM algorithm (WCEM) is obtained by introducing a standard C-step between the E-step and the weighted M-step. The main feature of the WCEM algorithm is that
 one single weight is attached to each unit, based on its current assignment after the C-step, rather than component-wise weights. Then, the resulting WLEE shows the same structure as in (\ref{wlee_mark}) but with the difference that $u_{ij}=1$ or $u_{ij}=0$.%, since it has been driven by a classification likelihood.
The WCEM is described as follows:
\begin{enumerate}
\item {\bf Initialization}: 
$$
\tau^{(0)}=(\pi^{(0)}, \mu_1^{(0)}, \ldots,  \mu_K^{(0)},  \Sigma_1^{(0)}, \ldots, \Sigma_K^{(0)}) \ .
$$

\item {\bf E step}:  
$$
u_{ik}^{(s)}=\frac{\pi_k^{(s-1)}\phi_p\left(y_i; \mu_k^{(s-1)},\Sigma_k^{(s-1)}\right)}{\sum_{k=1}^K \pi_k^{(s-1)}\phi_p\left(y_i; \mu_k^{(s-1)},\Sigma_k^{(s-1)}\right)}
$$

\item {\bf C step}:
let $k_i^{(s)}=\argmax_k u_{ik}^{(s)}$
identify the cluster assignment for the $i^{th}$ unit at the $s^{th}$ iteration. Then
$$
\tilde u_{ik}^{(s)}=\left\{
\begin{array}{cc}
1 & \textrm{if} \ k=k_i\\
0 &\textrm{if} \ k\neq k_i\\
\end{array}
\right. \ .
$$

\item {\bf Weighted M step}: based on current parameter estimates $\tau^{(s)}$ and cluster assignments $k_i$,
\begin{enumerate}
\item {\bf Soft-trimming}: evaluate the Mahalanobis-type distances of each point w.r.t. the component it belongs in
$$
d_{ik_i}^{(s)}=d\left(y_i; \mu_{k_i}^{(s-1)}, \Sigma_{k_i}^{(s-1)}\right) \ .
$$
Then,  compute the corresponding Pearson residuals and weights as
$$
\delta_{i}^{(s)}=\delta_{ik_i}^{(s)}=\frac{\hat m_n\left(d_{ik_i}^{(s)^2}\right)}{m_{\chi^2_p}\left(d^{(s)^2}_{ik_i}\right)}-1
$$
and 
$$
w_i^{(s)}=w_{ik_i}^{(s)}=\frac{\left[A\left(\delta_{ik_i}^{(s)}\right)+1\right]^+}{\delta_{ik_i}^{(s)}+1} \ 
$$
respectively, where 
%$$\hat m_n(d^2)=\frac{1}{n}\sum_{i=1}^n k(d^2; d^2_{ik_i}, h)=\frac{1}{n}\sum_{i=1}^n k(d^2; d^2_{ik}, h)\tilde u_{ik}.$$
%$$\hat m_n(d^2)=\frac{1}{\sum_{i=1}^n\tilde u_{ik}}\sum_{i=1}^n k(d^2; d^2_{ik}, h)\tilde u_{ik}.$$
\begin{eqnarray*}
\hat m_n\left(d_{ik_i}^{2}\right)&=&\frac{1}{\sum_{i=1}^n\tilde u_{ik_i}}k(d^2; d^2_{ik_i}, h)\\&=&\frac{1}{\sum_{i=1}^n\tilde u_{ik}} k(d^2; d^2_{ik}, h)\tilde u_{ik} \ .
\end{eqnarray*}
Hence, component-wise kernel density estimates only involve distances conditionally on cluster assignment.
%where $w_{ik}=0$ when $k\neq k_i$.

\item {\bf Update membership probabilities and component specific parameter estimates}: 
%\sum_{k=1}^K

\begin{eqnarray*}
\pi_k^{(s+1)}&=&\frac{\sum\limits_{i=1}^n \tilde u_{ik}^{(s)}w_{ik_i}^{(s)}}{\sum_{i=1}^n w_{ik_i}^{(s)}}\\
\mu_k^{(s+1)}&=&\frac{\sum\limits_{i=1}^n y_i w_{ik_i}^{(s)}\tilde u_{ik}^{(s)}}{\sum_{i=1}^n w_{ik_i}^{(s)}\tilde u_{ik}^{(s)}}\\
\Sigma_k^{(s+1)}&=&\frac{\sum\limits_{i=1}^n \left(y_i-\mu_k^{(s+1)}\right)\left(y_i-\mu_k^{(s+1)}\right)^{\T} w_{ik_i}^{(s)}\tilde u_{ik}^{(s)}}{\sum_{i=1}^n w_{ik_i}^{(s)}\tilde u_{ik}^{(s)}}.
\end{eqnarray*} 
\item {\bf Set} $\tau^{(s+1)}=\left(\pi^{(s+1)}, \mu_1^{(s+1)}, \ldots,  \mu_K^{(s+1)},  \Sigma_1^{(s+1)}, \ldots, \Sigma_K^{(s+1)}\right)$

\end{enumerate}

\end{enumerate}
It is worth noting that both the weighted algorithms returns weighted estimates of covariance. The final output can be suitably modified in order to provide unbiased weighted estimates.   

A strategy to select the best root is given in \cite{agostinelli2017weighted}.
%\cite{markatou1998} and \cite{agostinelli2006}. In particular, according to the latter proposal, 
The selected root is that with the lowest fitted probability
%For each starting point, Pearson residuals are evaluated and the solution  
\begin{equation}
\label{selection}
\Pr_{\hat\tau}\left[\delta(\hat d^2; \hat\tau, \hat M_n)<-0.95\right] \ .
\end{equation}
It is worth to stress that Pearson residuals involved in (\ref{selection}) have to be evaluated conditionally on cluster assignment based on the original multivariate data for purely selection purposes.
The sum of the weights also provides some guidance for root selection. Actually, if $\sum_{i=1}^n \hat w_i \approx 1$ than the WLE is close to the MLE, whereas if $\sum_{i=1}^n \hat w_i$ is too small, than the corresponding WLE is a degenerate solution, indicating that it only represents a small subset of the data. 
The fitted weights are evaluated w.r.t. the final cluster assignment, as well.

 \subsection{Eigen-ratio constraint}
It is well known that maximization of the mixture likelihood (\ref{mixlik}) or the classification likelihood (\ref{complik}) is an ill-posed problem since the objective function may be unbounded \citep{day1969estimating, maronna1974multivariate}. Therefore, 
in order to avoid such problems, the optimization is performed under suitable constraints.
In particular, we employed the eigen-ratio constraint defined as
\begin{equation}
\label{ER}
\frac{\max_j\max_k \lambda_j(\Sigma_k)}{\min_j\min_k \lambda_j(\Sigma_k)}\leq c, j=1,2\ldots,p, \ k=1,2,\ldots,K
\end{equation}
where $\lambda_j(\Sigma_k)$ denoted the $j^{th}$ eigenvalue of the covariance matrix $\Sigma_k$ and $c$ is a fixed constant not smaller than one aimed at tuning the strength of the constraint. For $c=1$ spherical clusters are imposed, while as $c$ increases varying shaped clusters are allowed.
The eigen-ratio constraint (\ref{ER}) can be satisfied at each iteration by adjusting the eigenvalues of each $\Sigma_k^{(s)}$. This is achieved by replacing them with a truncated version 
$$
\lambda_j^*(\Sigma_k)=\left\{
\begin{array}{cl}
c & \textrm{if} \ \lambda_j(\Sigma_k)<c\\
\lambda_j(\Sigma_k) &  \textrm{if} \ c\leq \lambda_j(\Sigma_k)\leq c\theta_c\\
c\theta_c &  \textrm{if} \ \lambda_j(\Sigma_k)> c\theta_c\\
\end{array}
\right.
$$
where $\theta_c$ is an unknown bound depending on $c$.
The reader is pointed to \cite{fritz2013fast, garcia2015avoiding} for a feasible solution to the problem of finding $\theta_c$.  \

\subsection{The selection of $h$}
The selection of $h$ is a crucial task. 
Let $\bar\omega=n^{-1}\sum_{i=1}^n w_i$ be the average of the weights at the true parameter value. Then, the expected value of $n(1-\bar \omega)$ tells us about the expected rate of contamination in the data. In the case of a normal kernel, \cite{markatou2000mixture} suggested to tune $h$ for a fixed (approximated) expected downweighting level. 
According to authors' experience (see \cite{agostinelli2017weighted}, \cite{agr_disc_sma} and \cite{greco2016weighted}, for instance), but also as already suggested by \cite{markatou1998}, a safe selection of $h$ can be achieved by monitoring the empirical downweighting level $(1-\hat{\bar \omega})$ as $h$ varies, with $\hat{\bar \omega}=n^{-1}\sum_{i=1}^n \hat w_i$, where the weights at convergence $\hat w_i$ are evaluated at the fitted parameter value, conditionally on the final cluster assignment.
The monitoring of WLE analyses has been applied successfully in \cite{agr_disc_sma} to the case of robust estimation of multivariate location and scatter. The reader is pointed to the paper by \cite{cerioli2018} for an account on the benefits of monitoring. 

%In particular, in the case of a mixture model fitted by the WEM or WCEM algorithm, the weights at convergence  $\hat w_i$ are meant to be evaluated conditionally on the final cluster assignment. 
The tuning of the smoothing parameter could be based on several quantities of interest stemming from the fitted mixture model. For instance,  one could monitor the weighted log-likelihood at convergence or unit-specific robust distances conditionally on the final cluster assignment,  rather than the empirical downweighting level. A good strategy would be to monitor different quantities, for a varying number of clusters.
For instance, if an abrupt change in the monitored empirical downweighting level or in the robust distances is observed, it may indicate the transition from a robust to a non robust fit and aid in the selection of a value of $h$ that gives an appropriate compromise between efficiency and robustness. 
It is worth to note that, the trimming level in {\tt tclust} or the improper density constant in {\tt otrimle} are selected in a monitoring fashion, as well.

\subsection{Classification and outlier detection}
%According to both algorithms, the value of $u_{ik}$ at convergence can be used for cluster assignment.
The WCEM automatically provides a classification of the sample units, since the value of $\tilde u_{ik}$ at convergence is either zero or one. With the WEM, by paralleling a common approach, a Maximum-A-Posteriori criterion can be used for cluster  assignment, that is a C-step is applied after the last E-step.
Such criteria lead to classify all the observations, both genuine and contaminated data,
meaning that also outliers are assigned to a cluster.
%, while receiving a final weight close to zero.  
Actually, we are not interested in classifying outliers and for purely clustering purposes outliers have to be discarded. 

We distinguish two main approaches to outlier detection. According to the first, outlier detection should be based on the robust fitted model and performed separately by using formal rules. The key ingredients in multivariate outlier detection are the robust distances \citep{rousseeuw1990unmasking, cerioli2010multivariate}. The reader is pointed to \cite{cerioli2011error} for a recent account on outlier detection. An observation is flagged as an outlier when its squared robust distance exceeds a fixed threshold, corresponding to the $(1-\alpha)-$level quantile of the reference (asymptotic) distribution of the squared robust distances. A common solution is represented by the use of the $\chi^2_p$ and popular choices are $\alpha=0.025$ and $\alpha=0.01$. In the case of finite mixtures, the main idea is that the outlyingness of each data point should be measured  conditionally on the final assignment. Hence, according to a proper testing strategy, an observation is declared as outlying when
\begin{equation}
\label{rule}
d_{ik_i}^2>\chi^2_{p;1-\alpha} \ , \ d_{ik_i}^2=(y_i-\hat\mu_{k_i})^{\T} \hat\Sigma_{k_i}(y_i-\hat\mu_{k_i}) \ .
\end{equation}

The second approach stems from hard trimming procedures, such as {\tt tclust}, {\tt rtclust} and {\tt otrimle}. These techniques are not meant to provide simultaneous robust fit and outlier detection based on formal testing rules, but outliers are identified with those data points falling in the trimmed set or assigned to the improper density component, respectively. Therefore, by paralleling what happens with hard trimming, one could flag as outliers those data points whose weight, conditionally on the final cluster assignment, is below a fixed (small) threshold. Values as 0.10 or 0.20 seem reasonable choices. Furthermore,
the empirical downweighting level represents a natural upper bound for the cut-off value, that would give an indication of the largest tolerable swamping and of the minimum feasible masking for the given level of smoothing.  This approach is motivated by the fact that the multivariate WLE shares important features with hard trimming procedures, even if it is based on soft trimming, as claimed in  \cite{agr_disc_sma}. %Actually, outliers are expected to exhibit small weights, that goes to zero as the Pearson residual function increases.

The process of outlier detection may result in type-I and type-II errors. In the former case, a genuine observation is wrongly flagged as outlier (swamping), in the latter case, a true outlier is not identified (masking). Swamped genuine observations are false positives, whereas masked outliers are false negatives.
%The choice of $\alpha$the threshold needs to be tuned in a fashion similar to the level $\alpha$ in the testing procedure.
According to the first strategy, the larger $\alpha$ the more swamping and the less masking. In a similar fashion, the higher the threshold the more swamping and the less masking will characterize the second approach to outlier detection. 

In the following, both approaches to outlier detection will be taken into account and critically compared.

\section{Illustrative examples}
\label{ill}
\subsection{Simulated data}
\begin{figure*}
\centering
\includegraphics[width=0.38\textwidth]{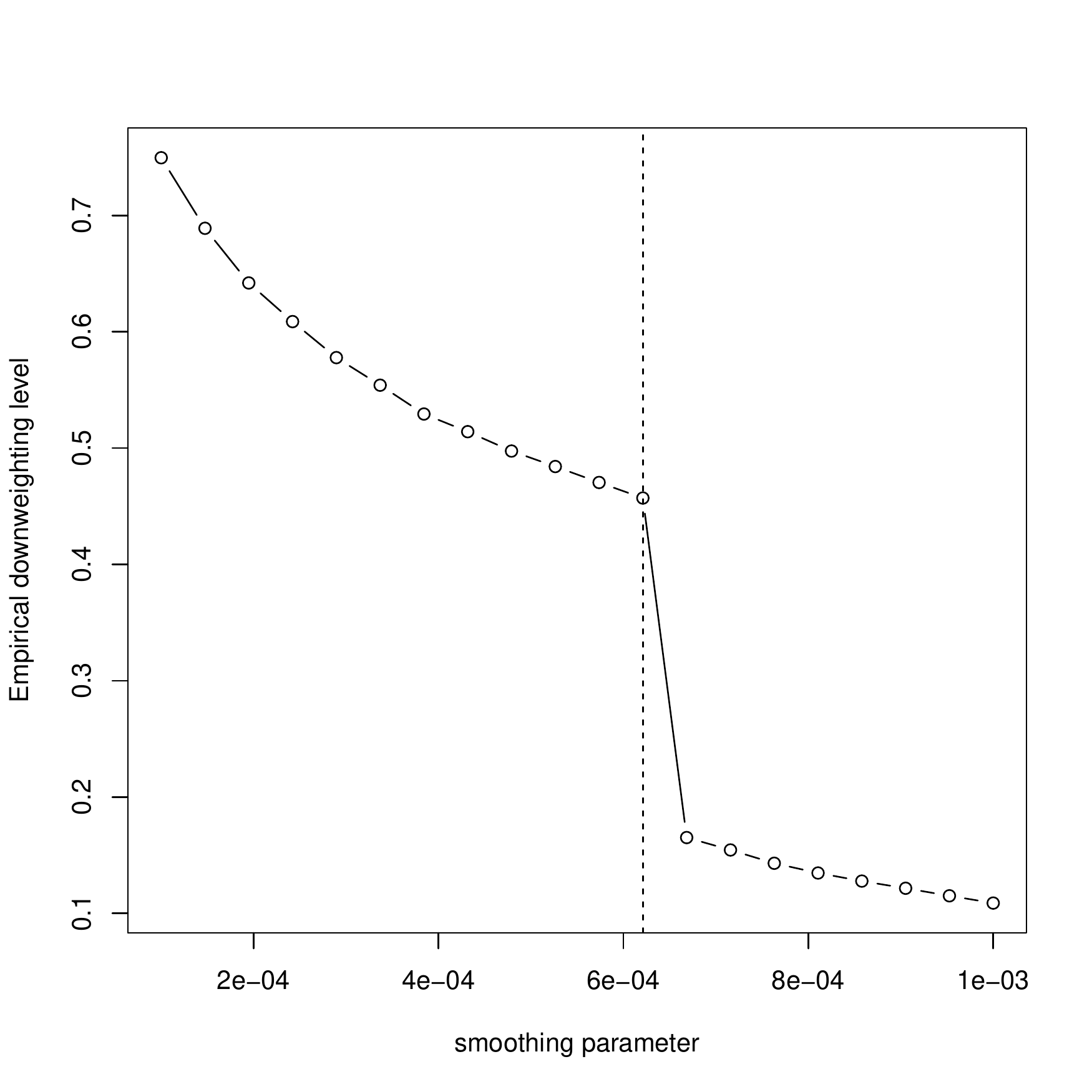}
\includegraphics[width=0.38\textwidth]{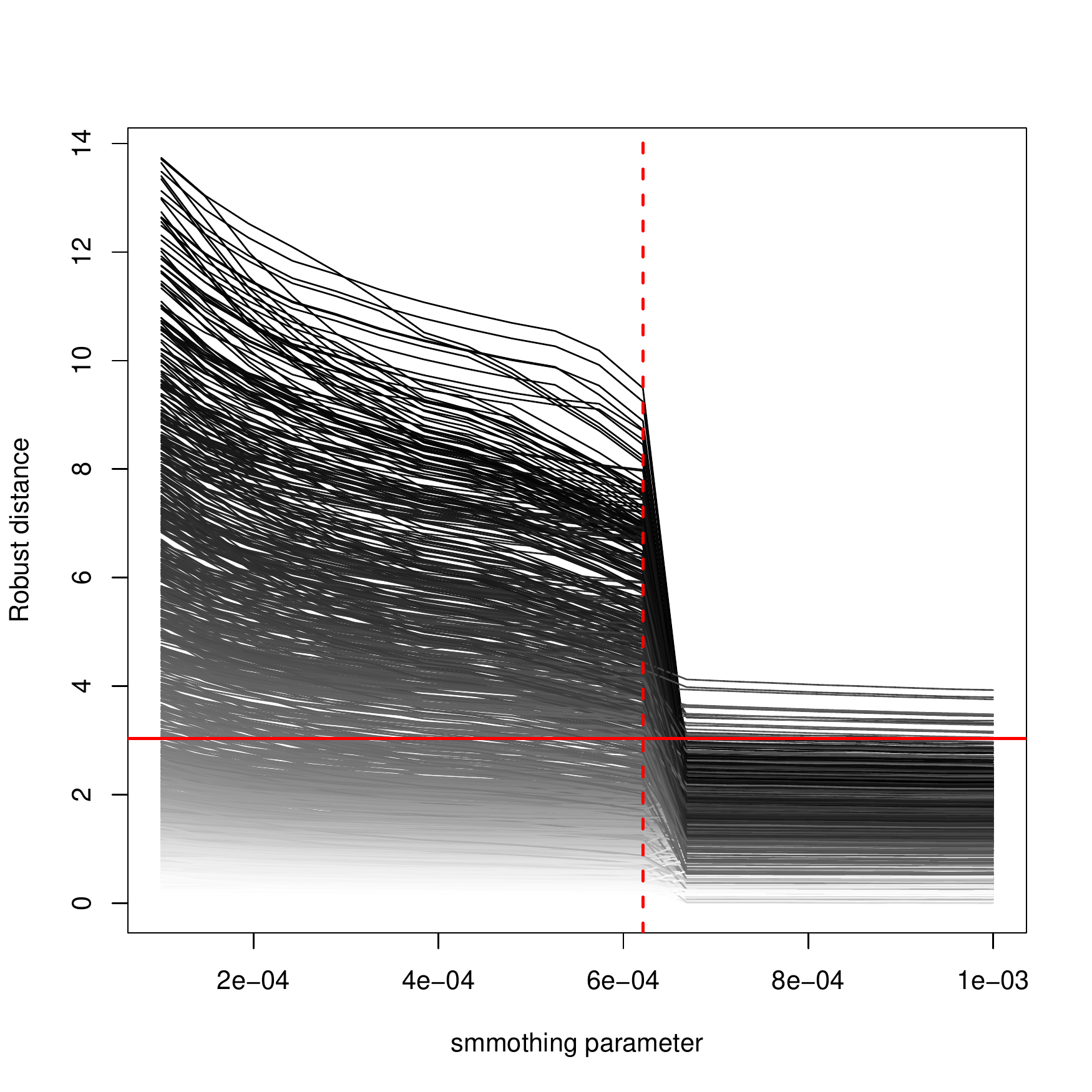}
\caption{Simulated data. Monitoring the empirical downweighting level (left) and robust distances (right) based on WEM. The vertical lines give the selected $h$. The horizontal line gives the $\chi^2_{2;0.99}$ quantile.}
\label{ese0}
\end{figure*}

\begin{figure*}
\centering
\includegraphics[width=0.38\textwidth]{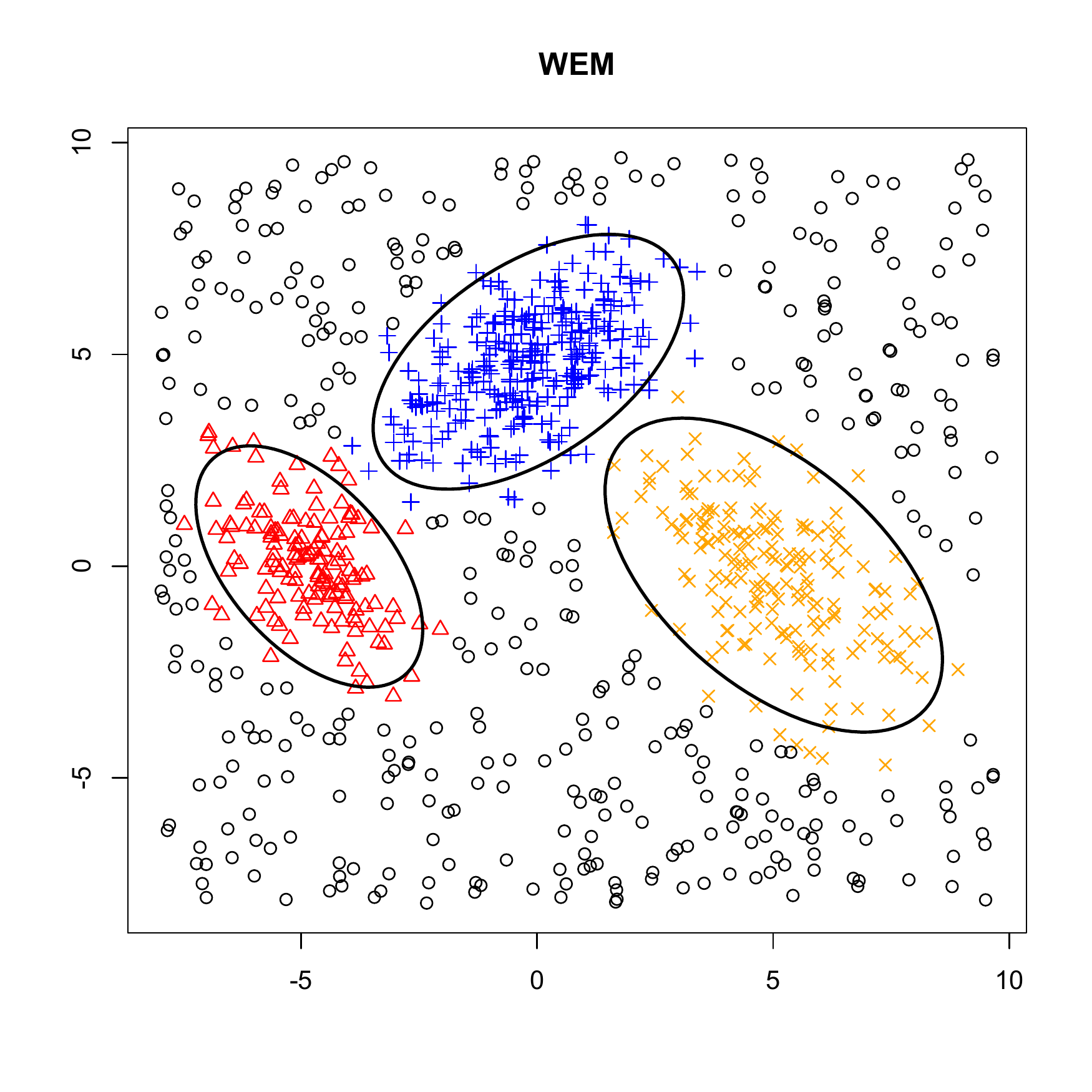}
\includegraphics[width=0.38\textwidth]{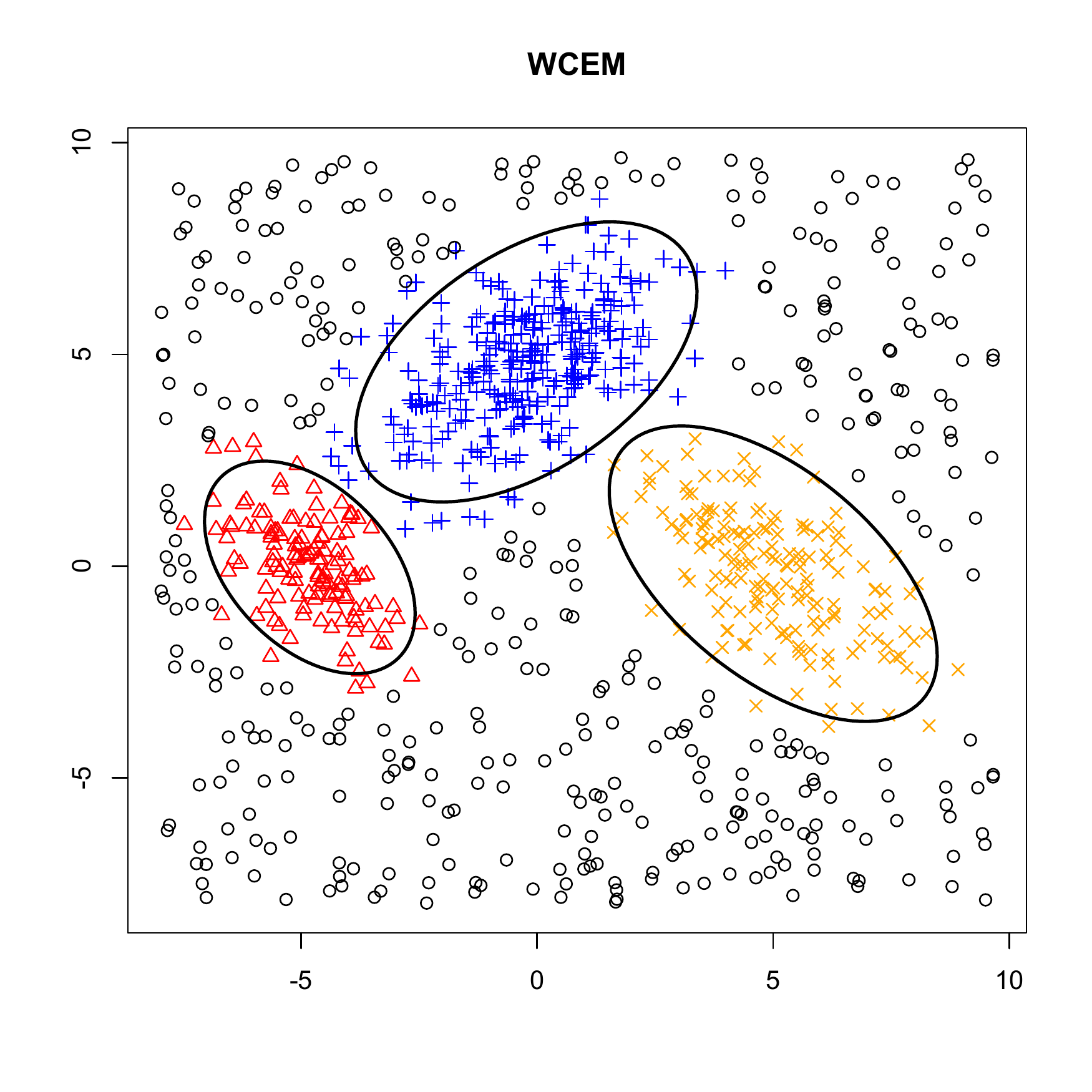}\\
\includegraphics[width=0.38\textwidth]{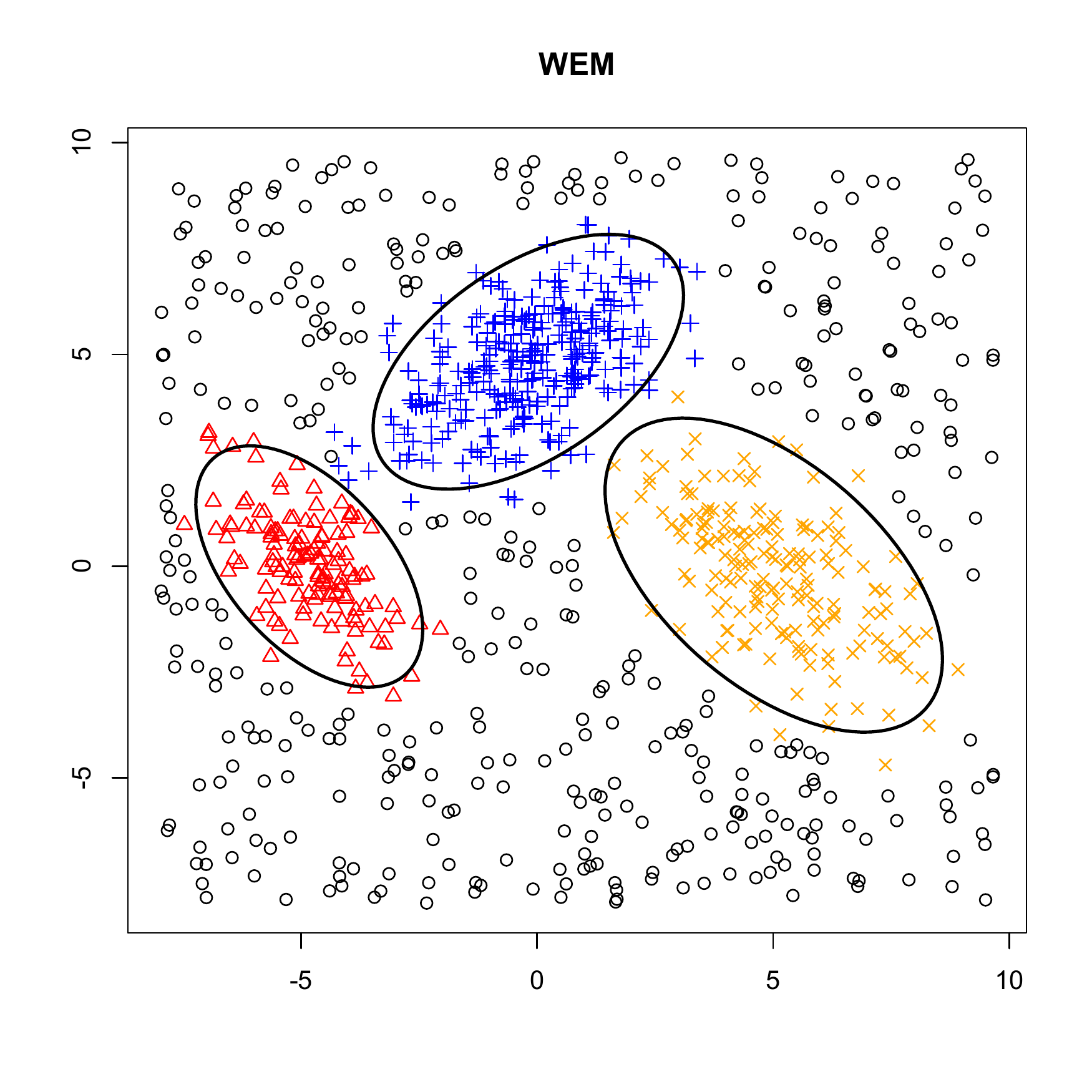}
\includegraphics[width=0.38\textwidth]{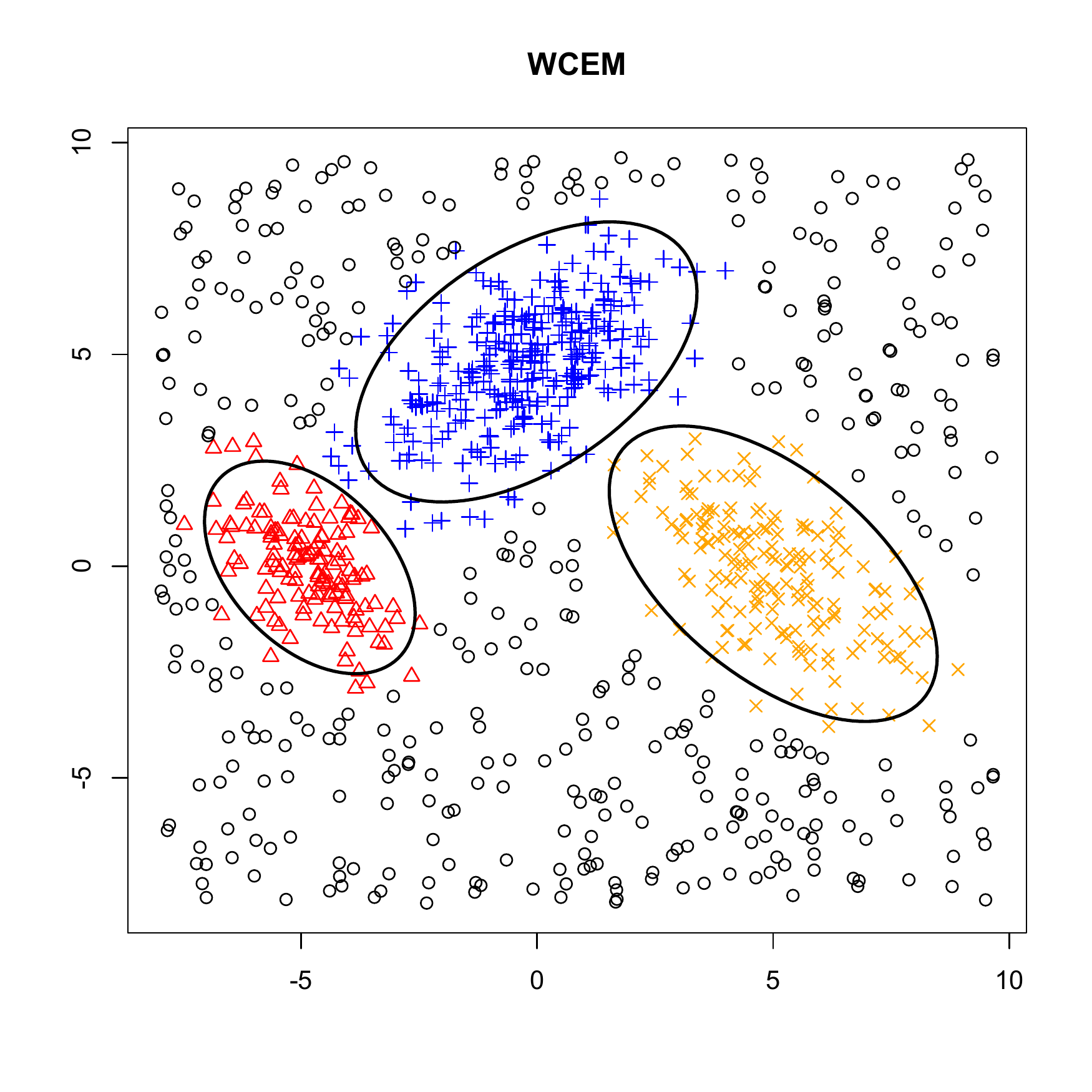}
\caption{Simulated data. Fitted components, cluster assignments and outlier detection by WEM (left) and WCEM (right). 
Top row: outlier detection based on $d^2_{k_i}<\chi^2_{p;0.99}$. Bottom row: outlier detection based on $w_{k_i}<0.2$
$95\%$ tolerance ellipses overimposed.}
\label{ese1}
\end{figure*}

Let us consider a three component mixture model with $\pi=(0.2,0.3,0.5)$, $\mu_1=(-5,0)^{\T}$, $\mu_2=(0,-5)^{\T}$, $\mu_3=(5,0)^{\T}$ and 
$$
\Sigma_1=\left(
\begin{array}{cc}
1&-0.5\\
-0.5 &1\\
\end{array}
\right), \quad
\Sigma_2=\left(
\begin{array}{cc}
2&1.25\\
1.25 &2\\
\end{array}
\right), \quad
\Sigma_3=\left(
\begin{array}{cc}
3&-1.75\\
-1.75 &3\\
\end{array}
\right).
$$
and a simulated sample of size $n=1000$, with $40\%$ of background noise. Outliers have been generated uniformly within an hypercube whose dimensions include the range of the data and are such that the distance to the closest component is larger than the $0.99$-level quantile of a $\chi^2_2$ distribution. 
WEM and WCEM have been run by setting the eigen-ratio restriction constant to $c=15$ (the true value is 9.5).
The weights are based on the Generalized Kullback-Libler divergence and a folded normal kernel.
The smoothing parameter $h$ has been selected by monitoring the empirical downweighting level and unit specific clustering-conditioned distances over a grid of $h$ values \citep{agr_disc_sma}. Figure \ref{ese0} displays the monitoring analyses of the empirical downweighting level and the robust distances for the WEM. In both panels an abrupt change is detected, meaning that for $h$ values on the right side of the vertical line the procedure is no more able to identify the outliers, hence being not robust w.r.t. the presence of contamination. Similar trajectories are observed for the WCEM. A color map has been used that goes from light gray to dark gray in order to highlight those trajectories corresponding to observations that are flagged as outlying for most of the monitoring. 
Figures \ref{ese1} displays the result of applying both the WEM and WCEM algorithm to the sample at hand with an outlier detection rule based on the 0.99-level quantile of the $\chi^2_2$ distribution and on a threshold for weights set at 0.2. Component-specific tolerance ellipses are based on the 0.95-level quantile of the $\chi^2_2$ distribution. 
We notice that both methods succeed in recovering the underlying structure of the clean data and that the outlier detection rules provide quite similar and satisfactory outcomes. The entries in Table \ref{t1_ese1} gives the rate of detected outliers $\epsilon$, swamping and masking stemming from the alternative strategies. 

\begin{table}
\caption{Simulated data. Outlier detection from different rules for WEM and WCEM.}
\label{t1_ese1}
%\centering
\begin{tabular}{r|r|ccc}
\hline\noalign{\smallskip}
&&$\epsilon$&swamp.&mask.\\
%\hline
%&\multicolumn{3}{c}{WEM}\\
\hline\noalign{\smallskip}
&$\alpha=0.010$&0.388&0.013&0.050\\
&$\alpha=0.025$&0.408&0.027&0.020\\
WEM&$\hat w<0.1$&0.353&0.003&0.123\\
&$\hat w<0.2$&0.386&0.013&0.055\\
&$\hat w<1-\hat{\bar w}$&0.429&0.052&0.005\\
\hline\noalign{\smallskip}
%&\multicolumn{3}{c}{WCEM}\\
%\hline
&$\alpha=0.010$&0.378&0.017&0.080\\
&$\alpha=0.025$&0.408&0.027&0.020\\
WCEM&$\hat w<0.1$&0.333&0.003&0.173\\
&$\hat w<0.2$&0.366&0.017&0.105\\
&$\hat w<1-\hat{\bar w}$&0.417&0.038&0.015\\
\hline\noalign{\smallskip}
\end{tabular}
\end{table}

\section{Model selection}
\label{mod}
In model based clustering, formal approaches to choose the number of components are based on the value of the log-likelihood function at convergence.  Criteria such as the BIC or the AIC are commonly used to select $K$ when running the classical EM algorithm. In a robust setting, in {\tt tclust} the number of clusters is chosen by monitoring the changes in the trimmed classification likelihood over different values of $K$ and contamination levels. A formal approach has not been investigated yet in the case of the {\tt otrimle}, even if the authors conjectures that a monitoring approach or the development of information criteria can be pursued as well.

Here, when the robust fit is achieved by the WEM algorithm, we suggest to resort to a weighted counterparts of the classical AIC or BIC criteria, according to the results stated in \cite{agostinelli2002robust}. Then, the proposed strategy is based on minimizing
\begin{equation}
Q^w(K)=-2\ell^w(y; \hat\tau)+m(K)
\label{wsel}
\end{equation}
where $\ell^w(y; \hat\tau)=\sum_{i=1}^n \hat w_{ik_i}\ell(y_i;\tau)$%, $\hat w_{ik_i}$ are unit specific weights evaluated after the last C-step, that is they are conditional on the final classification, 
and $m(K)$ is a penalty term reflecting model complexity. 

\begin{figure}
%\centering
\includegraphics[width=0.45\textwidth]{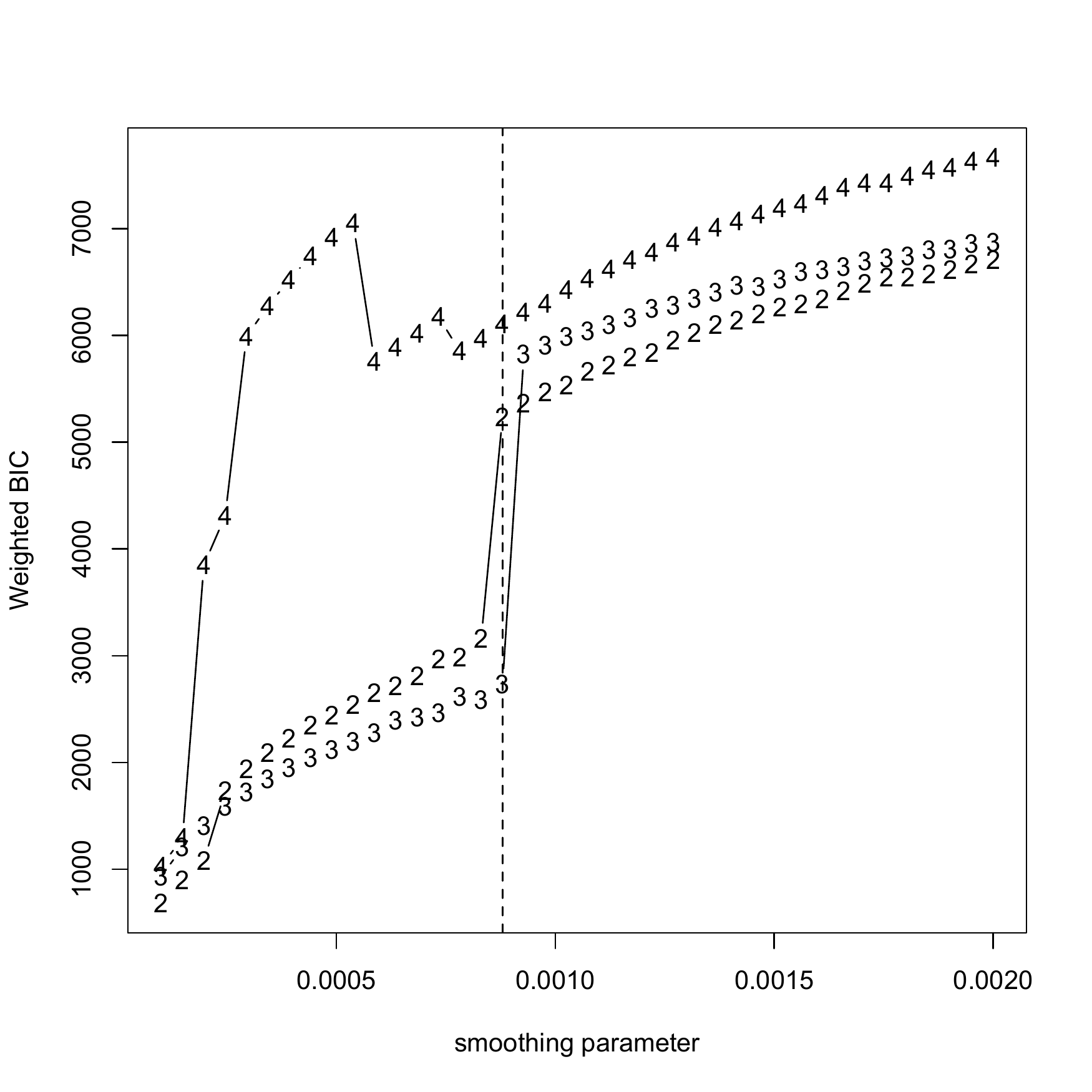}
\caption{Example 1. Monitoring the weighted BIC. The vertical line gives the selected $h$. }
\label{bic}
\end{figure}

Figure \ref{bic} displays the behavior of the weighted BIC for the set of simulated data of Section~\ref{ill}, for different choices of $K$ over a grid of values for the smoothing parameter $h$. A similar trajectory is observed for both $K=2$ and $K=3$. The abrupt change is detected at close values of $h$ but the choice $K=3$ is preferred since it leads to a smaller weighted BIC before the robust fit turns into a non robust one.
%Actually, one needs to keep in mind that different empirical downweighting levels can correspond to the same $h$ for different choices of $K$. Therefore, a fair strategy would be to compare $Q^w(K)$ values for similar values of $1-\bar w$.
 
Furthermore, for what concerns the WCEM algorithm, one could mimic the approach used in {\tt tclust} and monitor the weighted conditional likelihood at convergence for varying $K$ and $h$. Then, the number of clusters should be set equal to the minimum $K$ for which there is no substantial improvement in the objective function when adding one group more.

It can be proved that 
the robust criterion in (\ref{wsel}) is asymptotically equivalent to its classical counterparts at the assumed model, i.e. when the data are not prone to contamination. The proof is based on some regularity conditions about the kernel and the model that are required to assess the asymptotic behavior of the WLE  \citep{agostinelli2002robust, agostinelli2013weighted, agostinelli2017weighted}. In the case of finite mixture models, it is assumed further that an ideal clustering setting holds under the postulated mixture model, that is data are assumed to be well clustered. The following Theorem holds.\\

{\bf Theorem}. Let $\mathcal{Y}_j$ be the set of points belonging to the $j^{th}$ component, whose cardinality is $n_j$. The full data is defined as $\cup_{j=1}^K\mathcal{Y}_j$ with $\sum\limits_{j=1}^K=n$ and $\lim\limits_{n_j\rightarrow\infty}\frac{n_j}{n}=0$. Assume that (i) the model is correctly specified,  (ii) the WLE $\hat\tau$ is a consistent estimator of $\tau$, (iii) $\sup\limits_{y\in\mathcal{Y}_j} \left| w(\delta(y)) - 1 \right| \stackrel{p}{\longrightarrow} 0$. Then, $|Q^w(k)-Q(k)|\stackrel{p}{\rightarrow}0$.\\
%\begin{proof}
{\it Proof}. Let $\tilde\tau$ denote the maximum likelihood estimate. 
%We proceed by adding and removing the same quantity $\sum_i\ell(y_i;\hat\tau)$.
\begin{eqnarray*}
\frac{1}{2}|Q^w(k)-Q(k)|&=&\left|\sum_iw_i\ell(y_i;\hat\tau)-\sum_i\ell(y_i;\tilde\tau)\right|\\
&\leq&\left\{ \left|\sum_i (w_i-1)\ell(y_i;\hat\tau)\right|\right\}\\ &+&\left\{\left|\ell(y_i;\hat\tau)-\ell(y_i;\tilde\tau)\right|\right\}\\
&\leq&\sum_i|(w_i-1)\ell(y_i;\hat\tau)|\\
&\leq&\sup_y|w_i-1|\sum_i\ell(y_i;\hat\tau)\\
&\stackrel{p}{\rightarrow}0& \textrm{as} \ n_j\rightarrow\infty 
\end{eqnarray*}
%\end{proof}
%\smartqed
\section{Numerical studies}
\label{num}
In this section, we investigate the finite sample behavior of the proposed WEM and WCEM algorithms. 
Both algorithms are still based on non optimized {\tt R} code. Nevertheless, the results that follow are satisfactory and computational time always lays in a feasible range.  We set $n=1000$, $K=3$ and simulate data according to the {\it M5 scheme} as introduced in \cite{garcia2008general}. Clusters have been generated by $p$-variate Gaussian distributions with parameters
\begin{eqnarray*}
\mu_1&=&(-\beta,-\beta,0,\ldots,0), \\ \mu_2&=&(0,\beta,0,\ldots,0), \\ \mu_3&=&(\beta,0,0,\ldots,0)
\end{eqnarray*}
and
$$
\Sigma_1=\left(
\begin{array}{rrr}
15&-10&0_{p-2}\\
-10 &15&0_{p-2}\\
0_{p-2}^{\T}&0_{p-2}^{\T}&\mathrm{I}_{p-2}\\
\end{array}
\right), \ 
\Sigma_2=\mathrm{I}_p, \ 
\Sigma_3=\left(
\begin{array}{rrr}
45&0&0_{p-2}\\
0 &30&0_{p-2}\\
0_{p-2}^{\T}&0_{p-2}^{\T}&\mathrm{I}_{p-2}\\
\end{array}
\right),
$$
where $0_d$ is a null row vector of dimension $d$ and $\mathrm{I}_d$ is the $d\times d$ identity matrix.
Dimensions $p=2,5,10$ have been taken into account. The parameter $\beta$ regulates the degree of overlapping among clusters: smaller values yield severe overlapping whereas larger values give a better separation. Here, we set $\beta=6,8,10$. Theoretical cluster weights are fixed as $\pi=(0.2,0.4,0.4)$. 
Outliers have been generated uniformly within an hypercube whose dimensions include the range of the data and are such that the distance to the closest component is larger than the $0.99$-level quantile of a $\chi^2_p$ distribution. 
The rate of contamination has been set to $\epsilon=0.10,0.20$. 
The case $\epsilon=0$ has been used to evaluate the efficiency of the proposed techniques when applied to clean data.
The numerical studies are based on $500$ Monte Carlo trials. 
The weighted likelihood algorithms are both based on a folded normal kernel and a GKL RAF (with $\tau=0.9$), whereas we set $c=50$ as eigen-ratio constraint. The smoothing parameter $h$ has been selected in such a way that the empirical downweighting level lies in the range $(0.25, 0.35)$ under contamination, whereas it is about $10\%$ when no outliers occur. %Furthermore, outlier detection relies on the 0.99-level quantile of the $\chi^2_p$ distribution.
Fitting accuracy has been evaluated according to the following measures:
\begin{enumerate}
\item $||\hat\mu-\mu||$, where $\hat\mu$ and $\mu$ are $3\times p$ matrices with $\hat\mu_j$ and $\mu_j$ in each row, respectively, for $j=1,2,3$;
\item $\mathrm{ave_j}\log\mathrm{cond}\left(\hat\Sigma_j\Sigma_j^{-1} \right)$, where $\mathrm{cond}(A)$ denotes the condition number of the matrix $A$;
\item $||\hat\pi-\pi||$.
\end{enumerate}

For what concerns the task of outlier detection, several strategies have been compared: we considered a detection rule based on the 0.99-level quantile of the $\chi^2_p$ distribution, according to (\ref{rule}), but also based on the fitted weights, with thresholds set at $0.1$, $0.2$ and $1-\bar \omega$.
For each decision rule, for the contaminated scenario,
we report (a) the rate of detected outliers $\epsilon$; (b) the swamping rate; (c) the masking rate. The first is a measure of the fitted contamination level, whereas the others give insights on the level and power of the outlier detection procedure. We notice that comparisons across the different methods should be considered the more fair the closer  are the values of $\epsilon$.
For $\epsilon=0$, swamping only is taken into account.

Classification accuracy has been measured by (i) the Rand index and (ii) the misclassification error rate (MCE), both evaluated over true negatives for the robust techniques. %The result are based on the testing decision rule (\ref{rule}).
In order to avoid problems due to label switching issues, cluster labels have been sorted according to the first entry of the fitted location vectors. 

Under the assumed model, WEM and WCEM have been initialized by {\tt tclust} with $20\%$ of trimming and their behavior have been compared with the EM and CEM algorithms and the {\tt otrimle}, for the same eigen ratio constraint and the same initial values.
In the presence of contamination, we do not report the results concerning the non robust EM and CEM but only those regarding the WEM, WCEM, {\tt otrimle}  and {\it oracle} {\tt tclust}, i.e. with trimming level equal to the actual contamination level ({\tt tclust10} and {\tt tclust20}). In this case, starting values have been driven by 
{\tt tclust} with $50\%$ of trimming. 

It is worth to stress, here, that the comparison in terms of outlier detection reliability between weighted likelihood estimation,  {\tt tclust} and {\tt otrimle} can be considered fair 
only by looking at  the rate of weights below the fixed threshold for the former methodology and
trimmed observation or those assigned to the improper density group for the latter techniques,
%for the task of outlier detection, 
since formal testing rules have not ben considered neither for {\tt tclust} or {\tt otrimle}.

First, let us consider the behavior of WEM and WCEM at the assumed model. The entries in Table \ref{tab1no} give the considered average measures of fitting accuracy;  Table \ref{tab2no} gives the level of swamping according to the different strategies for WEM and WCEM, that are based on the $\chi^2_p$ distribution and the inspection of weights; Table \ref{tab3no} reports classification accuracy. 
The overall behavior of WEM and WCEM is appreciable: we observe a tolerable efficiency loss, a negligible swamping effect and a reliable classification accuracy, indeed, as compared with the non robust procedures. Furthermore, the results are quite similar to those stemming from {\tt otrimle} and quite often the inspection of the weights from WEM and WCEM leads to a smaller number of false positives, on average. 

%, that correspond to increasing levels of overlapping, as tuned by the parameter $\beta$
The performance of WEM and WCEM under contamination is explored next.
%The tables correspond to increasing levels of overlapping, as tuned by the parameter $\beta$, both in the uncontaminated and contaminated scenario.
The fitting accuracy provided by the proposed weighted likelihood based strategies is illustrated in Tables \ref{tab1}, \ref{tab2}, \ref{tab3}. In all considered scenarios, the behavior of WEM and WCEM is satisfactory and they both compare well with the oracle {\tt tclust} and {\tt otrimle}. 
In particular, the good performance of WEM and WCEM has to be remarked in the challenging situation of severe overlapping. Furthermore, for all data configurations, we notice the ability of WEM to combine accurate estimates of component-specific parameters with those of the cluster weights. 
The entries in Tables \ref{tab7}, \ref{tab8}, \ref{tab9} show the behavior of the testing procedure based on the $\chi^2_p$ distribution and the inspection of weights for all considered scenarios.  The empirical level of contamination is always larger than the nominal one but it is acceptable and stable as $p$ and $\beta$ change. Masking is always negligible hence highlighting the appreciable power of the testing procedure. We remark that one could also consider multiple testing adjustments in outlier detection as outlined in \cite{cerioli2011error}.
To conclude the analysis, Tables \ref{tab4}, \ref{tab5}, \ref{tab6} give the considered measures of classification accuracy as $\beta$ varies.
The results are quite stable across the four methods and all dimensions. As well as before, WEM and WCEM lead to a satisfactory classification, even in the challenging case of severe overlapping.

\section{Real data examples}
\label{real}
\subsection{Swiss bank note data}
\begin{figure}[htb]
%\centering
\includegraphics[width=0.45\textwidth]{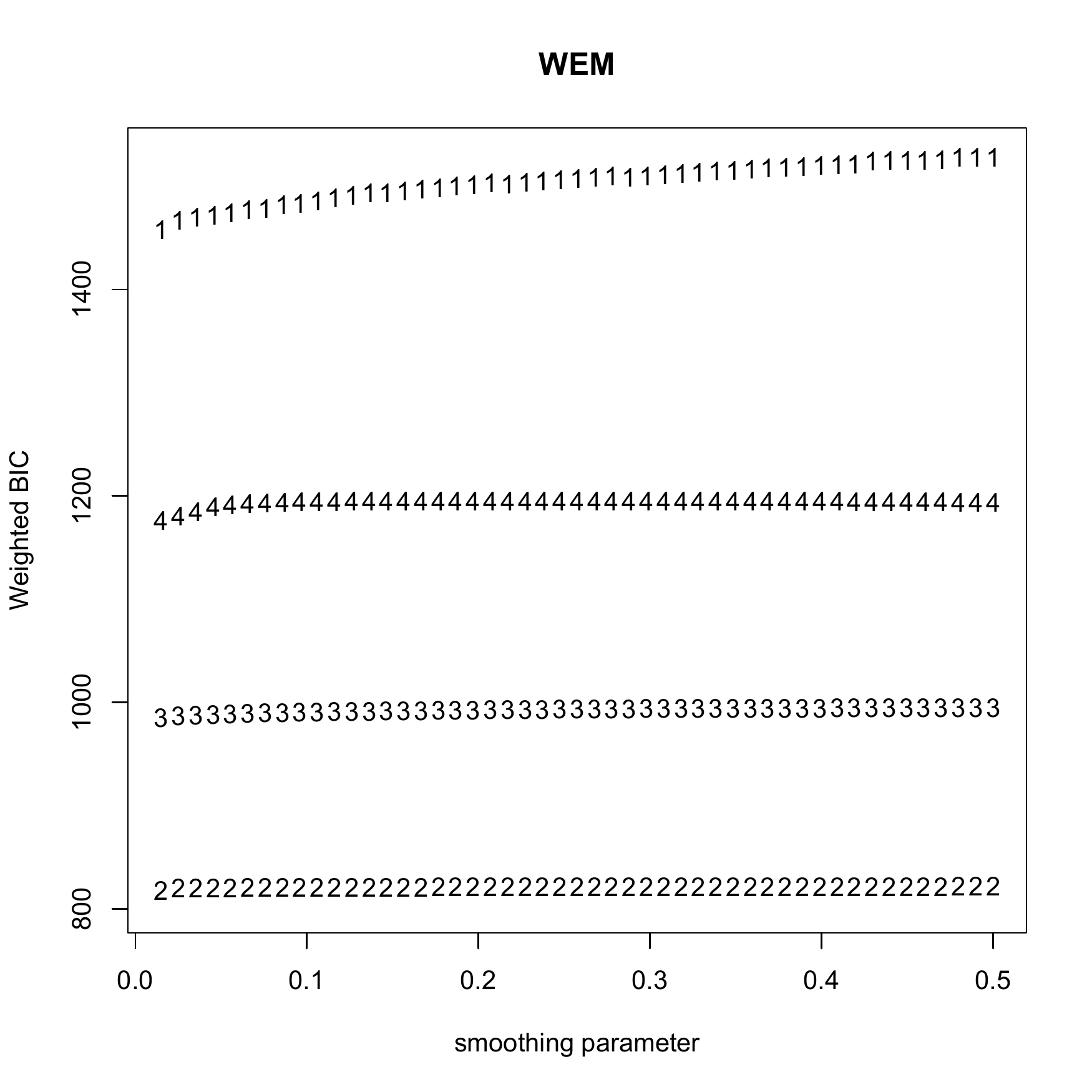}
\caption{Swiss banknote data. Monitoring the weighted BIC for WEM,  $K=1, 2, 3, 4$.}
\label{bic_swiss}
\end{figure}

\begin{figure*}[htb]
\centering
\includegraphics[width=0.75\textwidth]{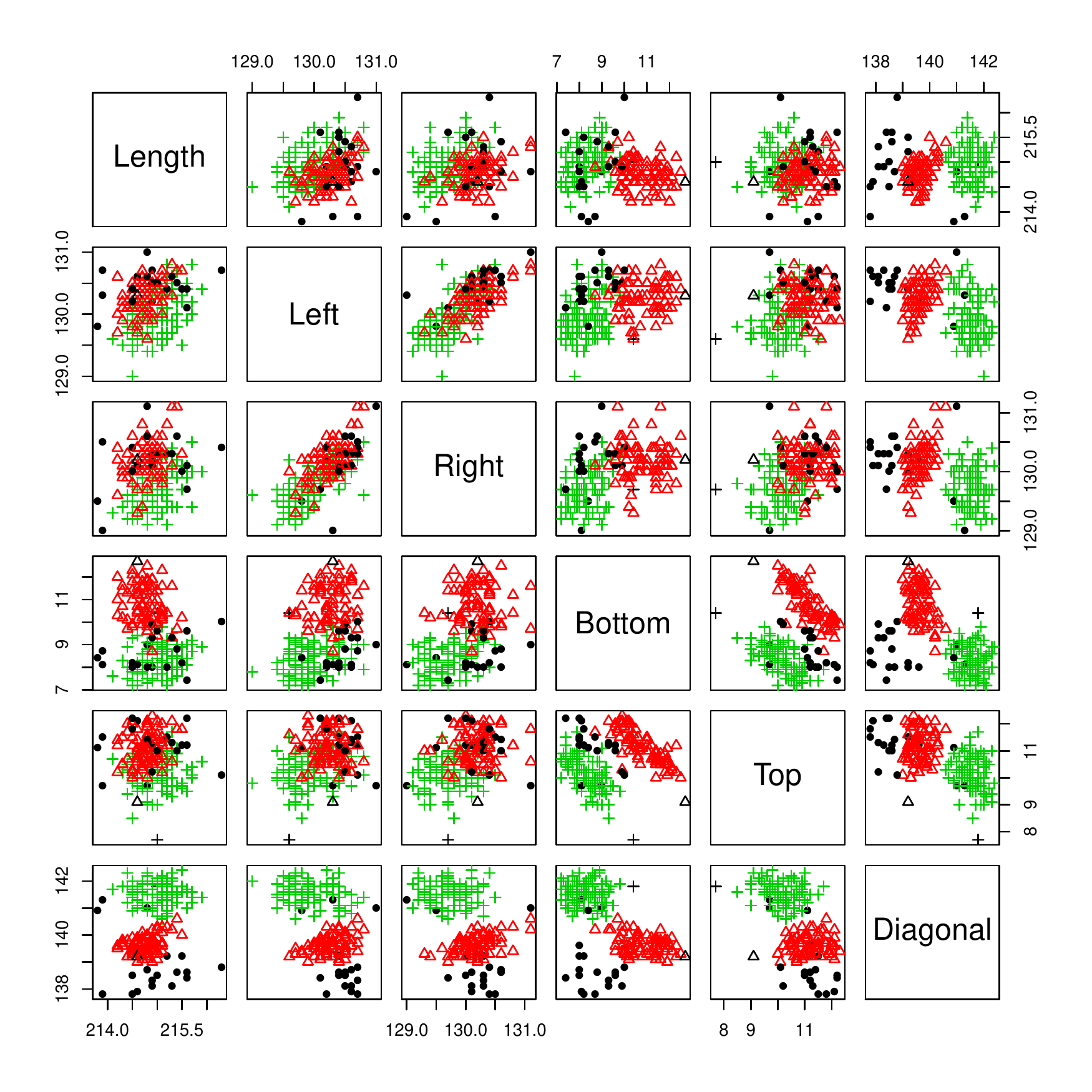}
\caption{Swiss banknote data. Cluster assignments by WEM. Observation whose weight is lower than $1-\bar w$ are considered outliers. Genuine bills are denoted by a green $+$, forged bills by a red $\triangle$. Outliers are denoted by a black filled circle.}
\label{swiss_fit}
\end{figure*}

\begin{figure*}[htb]
\centering
\includegraphics[width=0.38\textwidth]{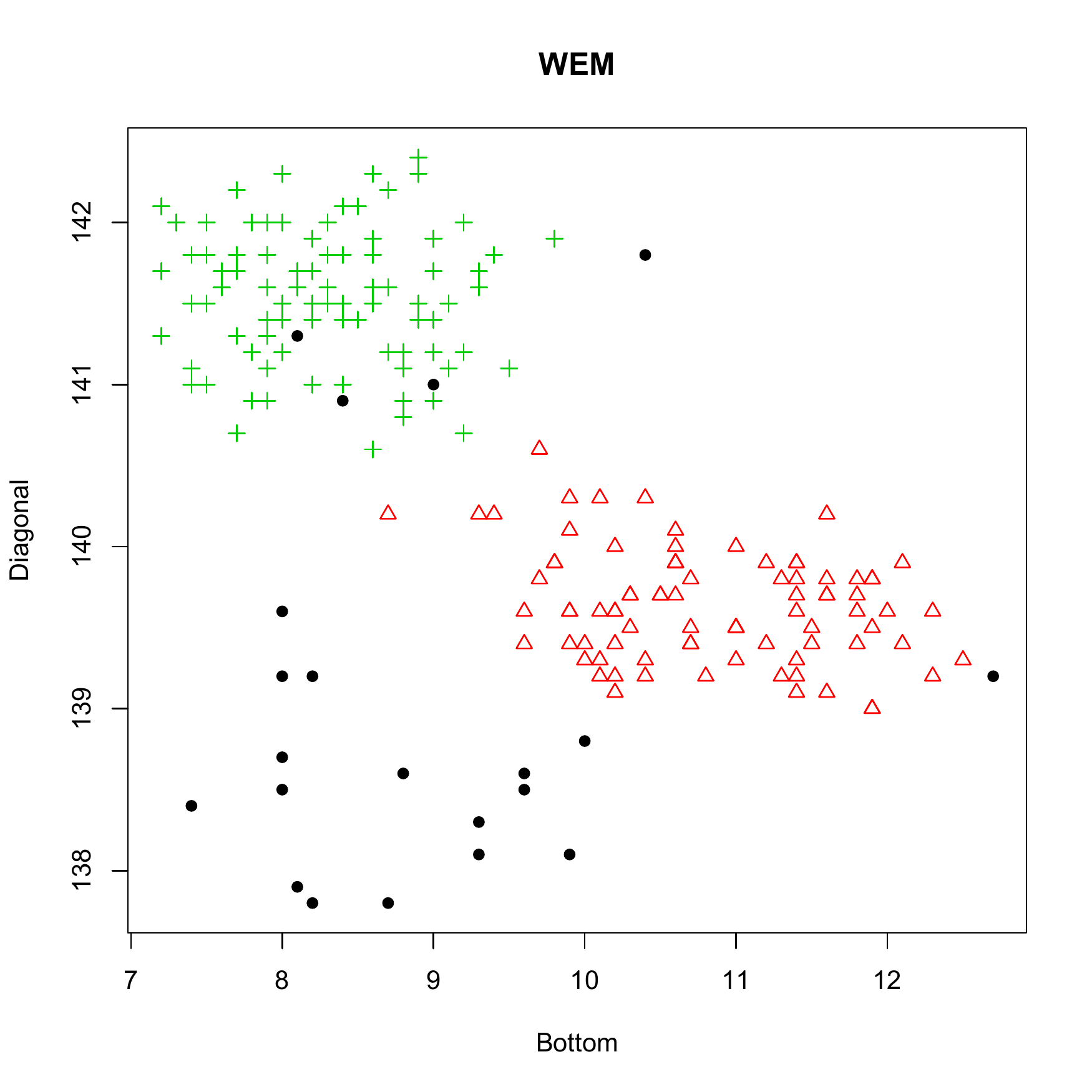}
\includegraphics[width=0.38\textwidth]{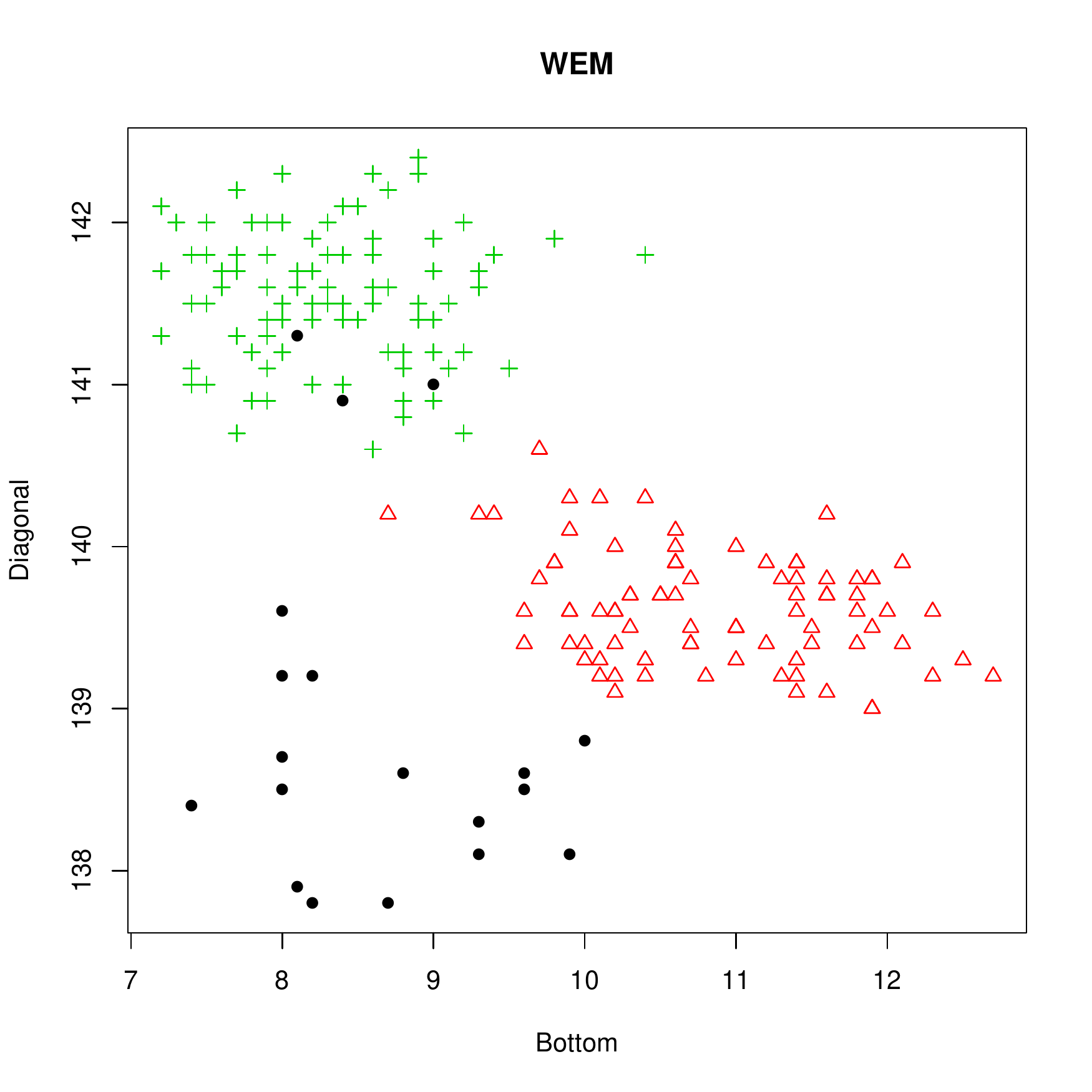}\\
\includegraphics[width=0.38\textwidth]{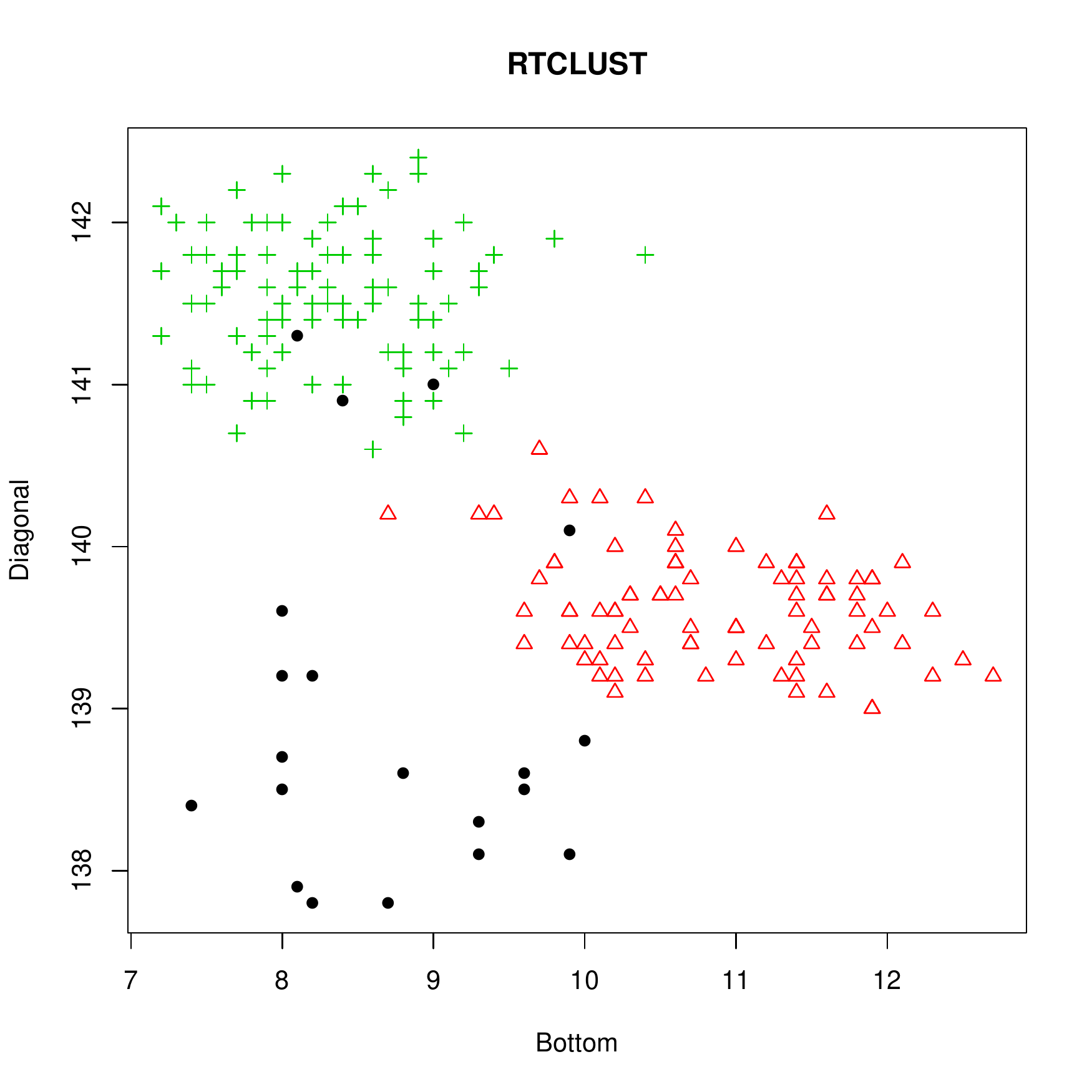}
\includegraphics[width=0.38\textwidth]{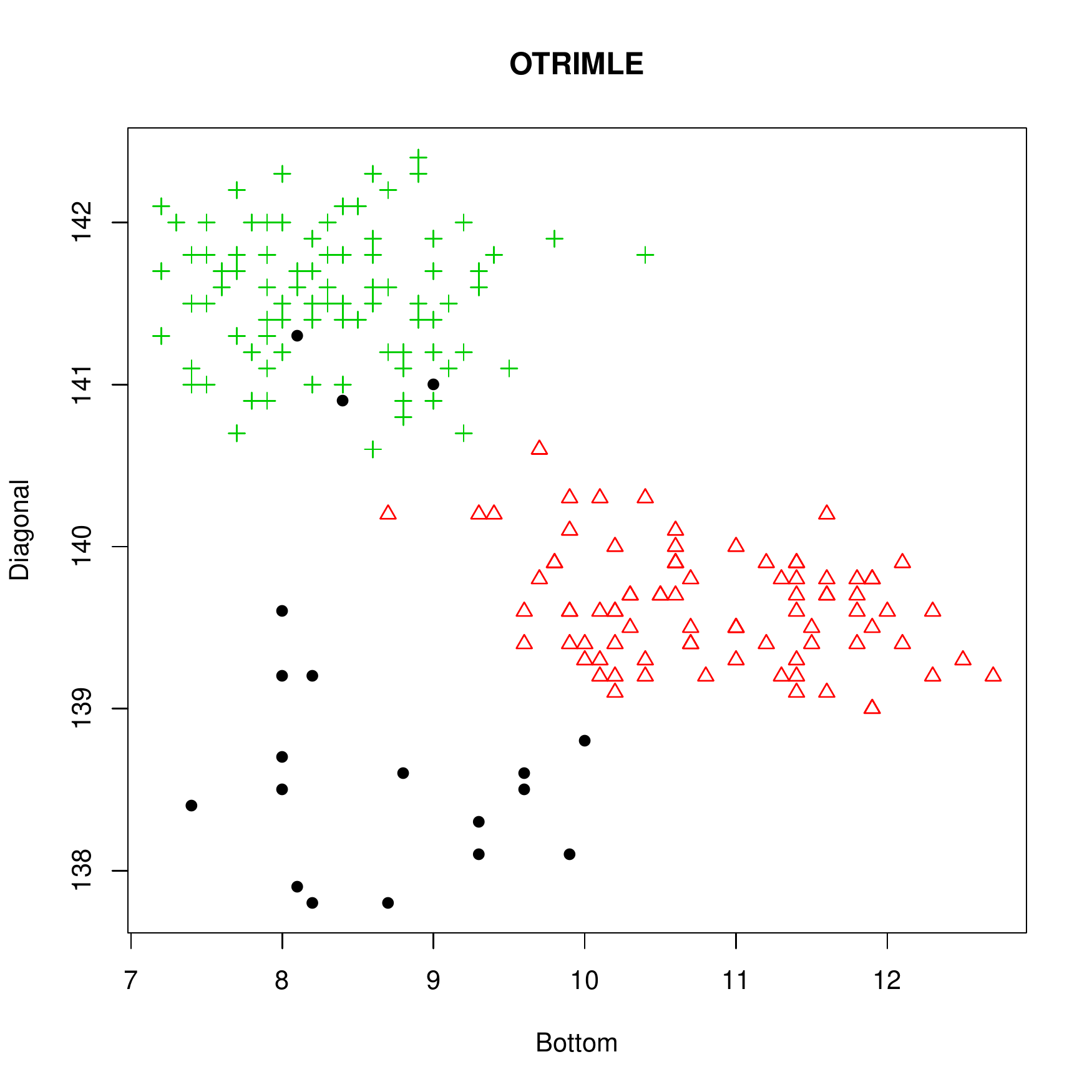}
\caption{Swiss banknote data. Fourth against the sixth variable with cluster assignments by WEM ($\alpha=0.01)$, WEM ($w_{k_i}<1-\bar w)$, {\tt rtclust} and {\tt otrimle} in clockwise fashion. Genuine bills are denoted by a green $+$, forged bills by a red $\triangle$. Outliers are denoted by a black filled circle.}
\label{all}
\end{figure*}

Let us consider the well known Swiss banknote dataset concerning $p=6$ measurements of $n=200$ old Swiss 1000-franc banknotes, half of which are counterfeit. 
The weighted likelihood strategy is based on a gamma kernel and a symmetric Chi-square RAF.
Our first task is to choose the number of clusters. To this end we look at the weighted BIC (\ref{wsel}) on a fixed grid of $h$ values for $K=1,2,3,4$ and a restriction factor $c=12$.
The inspection of Figure \ref{bic_swiss} clearly suggests a two groups structure for all considered values of the smoothing parameter $h$. %A similar plot is obtained for the WCEM algorithm. 
%Then, we turn into the problem of selecting the smoothing parameter $h$
The empirical downweighting level is fairly stable for a wide range of $h$ values. We decided to set $h=0.05$ leading to an empirical downweighting level equal to $0.10$.
%The algorithm has been initialized from {\tt tclust50}.
%chosen by looking at the $G-$statistic in Figure \ref{swiss_G}, that lead to select $h=0.033$ and $h=0.027$, respectively. 
The WEM algorithm based on the testing rule (\ref{rule}) with $\alpha=0.01$ leads to identify 21 outliers, that include 15 forged and 6 genuine bills.  
On the contrary, 
there are 19 data points whose weight is lower than $1-\hat{\bar w}$, that include 14 forged and 5 genuine bills.The cluster assignments stemming from the latter approach is displayed in Figure \ref{swiss_fit}. It is worth to note that the outlying forged bills coincide with the group that has been recognized to follow a different forgery pattern and characterized by a peculiar length of the diagonal (see \cite{garcia2011exploring, dotto2016reweighting} and references therein). On the other hand, the outlying genuine bills all exhibit some extreme measures. 
For the same value of the eigen-ratio constraint, the {\tt otrimle} assignes 19 bills to the improper component density, leading to the same classification of the WEM, whereas 
{\tt rtclust} includes in the trimming set one counterfeit bill more.
% the latter classifies one genuine bill less to the improper component density. 
A visual comparison between the three results is possible from Figure \ref{all}, whose panels
show a scatterplot of the fourth against the sixth variable with the classification resulting from WEM (with both outlier detection rules), {\tt rtclust} and {\tt otrimle}, respectively.
%The same result stems from WCEM.
The WCEM has been tuned to achieve the same empirical downweighting level and leads to the same results. 

\subsection{2018 World Happiness Report data}
In this section the weighted likelihood methodology is applied to a dataset from the 2018 World Happiness Report by the United Nations Sustainable Development Solutions Network \citep{happy} (hereafter denoted by WHR18). The data give measures about six 
key variables used to explain the variation of  subjective well-being across countries: per capita Gross Domestic Product (on log scale), Social Support,
i.e. the national average of the binary responses to the Gallup World Poll (GWP) question {\it If you were in trouble, do you have relatives or friends you can count on to help you whenever you need them, or not?},
 Health Life Expectancy at birth, Freedom to make life choices, i.e. the national average of binary responses to the GWP question {\it Are you satisfied or not with your freedom to choose what you do with your life?}, Generosity, measured by the residual of regressing the national average of GWP responses to the question {\it Have you donated money to a charity in the past month?} on GDP per capita, perception of Corruption, i.e. the national average of binary responses to the GWP questions {\it Is corruption widespread throughout the government or not?} and {\it Is corruption widespread within businesses or not?}.
The dataset is made of 142 rows, after the removal of some countries characterized by missing values.
The objective is to obtain groups of countries with a similar behavior, to identify possible countries with anomalous and unexpected traits and to highlight those features that are the main source of separation among clusters. 

In this example, a GKL RAF has been chosen, the unbiased at the boundary kernel density estimate  has been obtain by first evaluating a kernel density estimate on the log-transformed squared distance over the whole real line and then back-transforming the fitted density  to $(0, \infty)$ \citep{agostinelli2017weighted},  
and we set $c=50$.

\begin{figure*}
\centering
\includegraphics[width=0.38\textwidth]{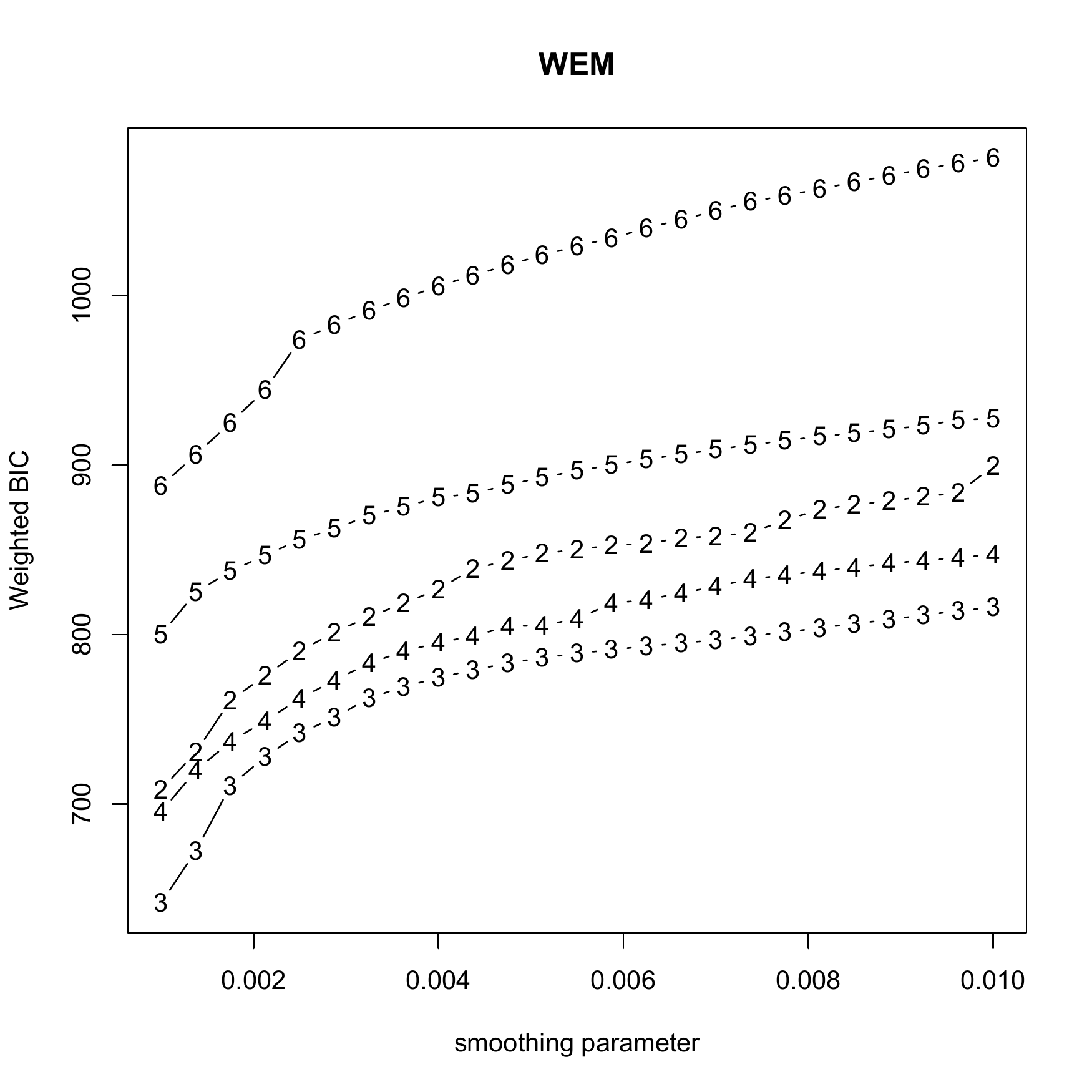}
\includegraphics[width=0.38\textwidth]{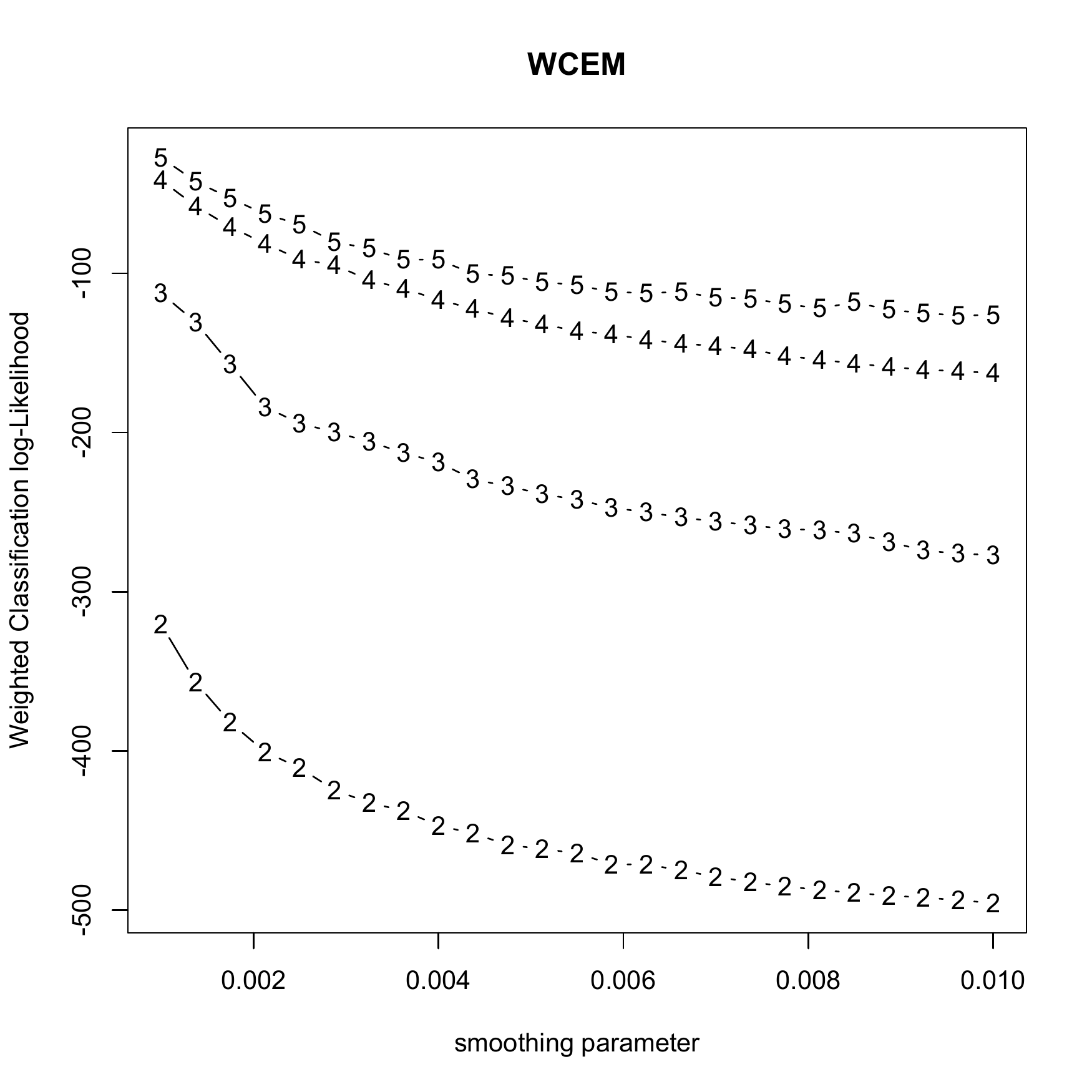}
\caption{WHR18 data. Monitoring the weighted BIC for WEM (left) and the weighted Classification log-Likelihood for WCEM, $K=2, 3, 4, 5, 6$.}
\label{whr_wbic}
\end{figure*}

\begin{figure}
%\centering
\includegraphics[width=0.45\textwidth]{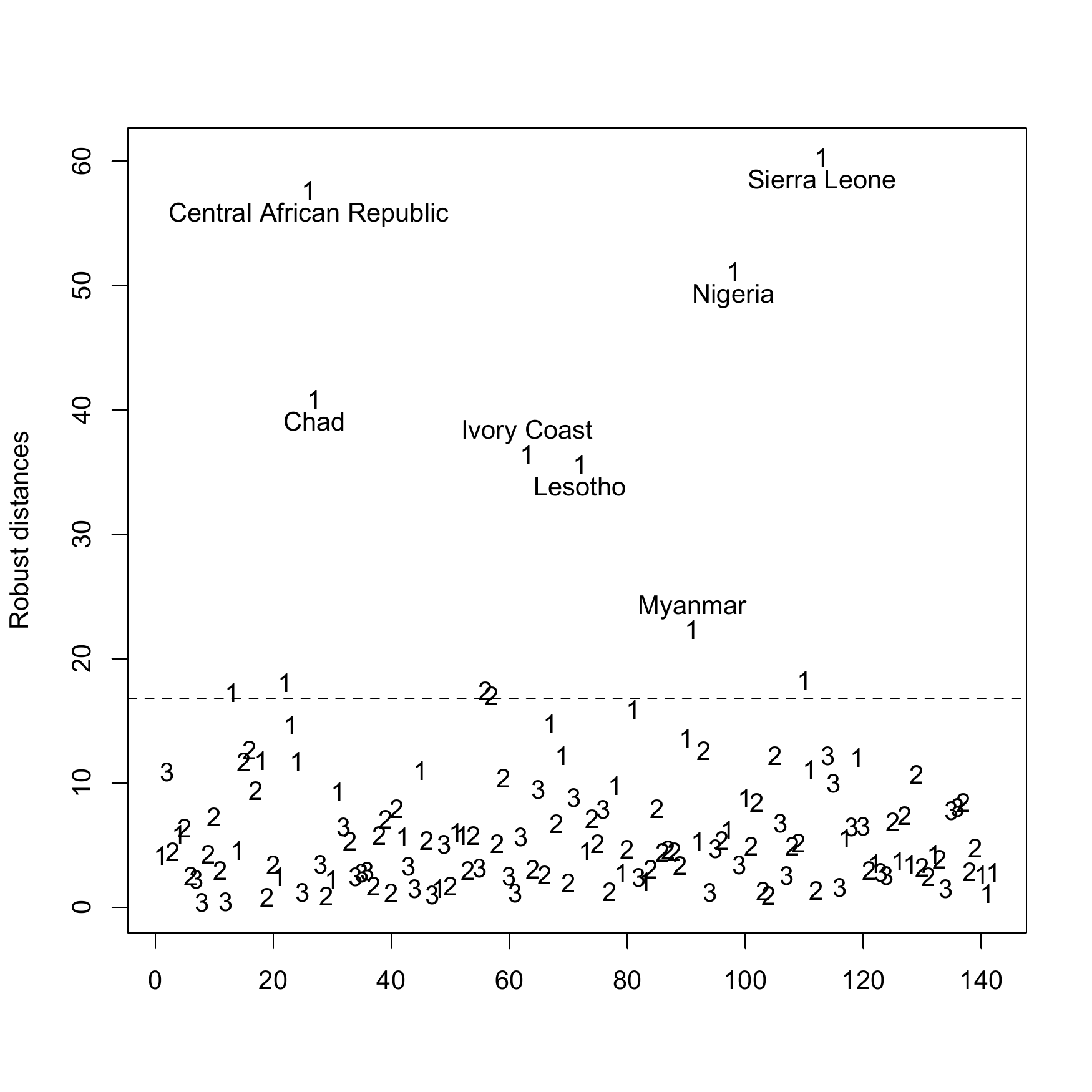}
\caption{WHR18 data. Distance plot for WEM with $K=3$. The horizontal line gives the $\chi^2_{6;0.99}$ quantile.}
\label{whr_distplot}
\end{figure}

In order to select the number of clusters, we monitored the weighted BIC stemming from WEM and the Classification log-Likelihood at convergence from WCEM for different values of $K$ and $h$. The corresponding monitoring plots are given in Figure \ref{whr_wbic}, respectively. Based on the WEM algorithm, $K=3$ is to be preferred, even if the gap with the case $K=4$ is very small for all considered values of the smoothing parameter. 
%Moreover, both choices of $K$ lead to almost the same empirical downweighting level for all $h$ values.
On the other hand, the inspection of the weighted Classification log-Likelihood driven by the WCEM suggests $K=4$. Therefore, we have applied our WEM and WCEM algorithms both based on $K=3$ and $K=4$. As with $K=4$ two groups are not very separated, we preferred $K=3$ and reported only those results for reasons of space.
Moreover, the results stemming from WEM and WCEM were very similar both in terms of fitted parameters, cluster assignments and detected outliers. Then, in the following we only give the results driven by WEM. 
The empirical downweighting level was found not to depend in a remarkable fashion on the number of groups. In particular, for $K=3$, in the monitoring process of $1-\hat{\bar w}$ we did not observe any abrupt change but a smooth decline until a stabilization of the level of contamination occurred.
Then, we decided to use a $h$ value leading to $(1-\hat{\bar w})\approx 0.10$.
Figure~\ref{whr_distplot} displays the distance plot stemming from WEM. According to (\ref{rule}) for a level $\alpha=0.01$, 12 outliers are detected. A closer inspection of the plot unveils that some of such points are close to the cut-off value. Therefore, they are not considered as outliers but correctly assigned to the corresponding cluster. Furthermore, we notice that all the points leading to the largest distances are attached a very small weight ($<0.01$). The weight corresponding to Myanmar is about 0.40. The other countries near the threshold line all show weights about equal to 0.80.

The cluster profiles
and raw measurements for the detected outlying countries are reported in Table \ref{tab_whr}.
The three clusters are well separated in terms of all of the items considered, 
even if small differences are seen in terms of perception of Corruption between clusters 1 and 2.
Cluster 3 includes all the countries characterized by the highest level of subjective well-being. The differences with the other two clusters concerning GDP, HLE and Corruption are outstanding. On the opposite, in cluster 1 we find all the countries with the hardest economic and social conditions. 
For what concerns the explanation of the outliers, we notice that the subgroup of African countries composed by Central African Republic, Chad, Ivory Coast, Lesotho, Nigeria and Sierra Leone might belong to group 1 but is characterized by the six lowest HLE values. %Other countries such as Belize, India, Indocina, Myanmar and Rwanda might belong to cluster 1 or 2 but they all show high levels of Freedom that are in complete disagreement with them. 
For what concerns Myanmar, it exhibits extremely large Freedom and Generosity indexes and a surprising small value for Corruption. It should belong to cluster 1 but it is closer to cluster 3 according to the last three measurements, indeed.
For instance, it may be supposed that such measurements are not completely reliable because of problems with the questionnaires and the sampling or they may revel a surprising positive attitude despite of the difficult economic and life conditions. 

\begin{figure*}
\centering
\includegraphics[width=0.75\textwidth]{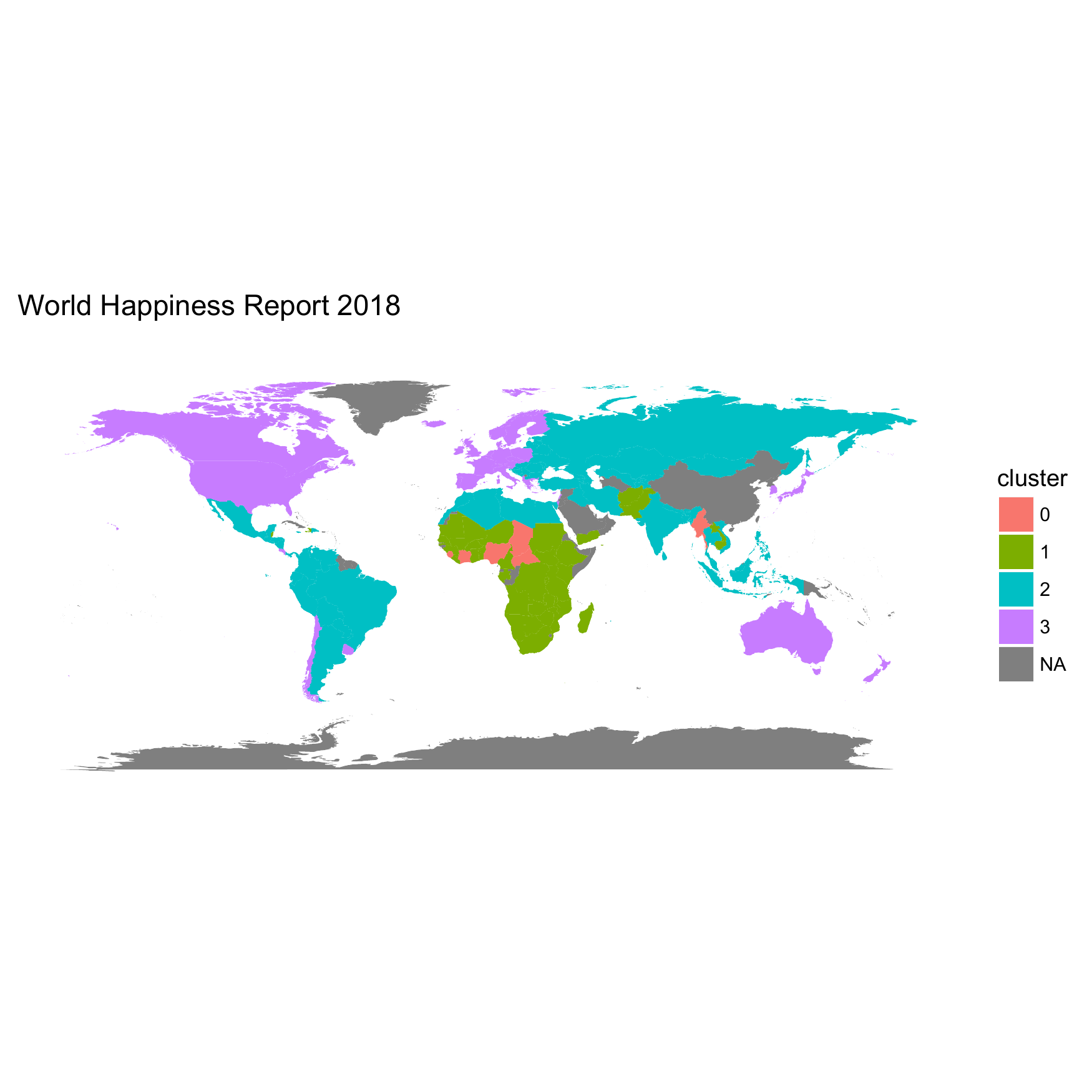}
\caption{WHR18 data. Spatial classification from WEM with $K=3$.}
\label{map}
\end{figure*}

A spatial map of cluster assignments is given in Figure \ref{map}, that confirms the goodness and coherence of the results and the ability of the considered six features to find reasonable clusters. Cluster 1 is mainly localized in Africa, cluster 2 is composed by developing countries, whereas cluster 3 includes the world leading countries, among which there are USA, Canada, Australia and the countries in the European Union.

\section{Conclusions}
We have proposed a robust technique for fitting a finite mixture of multivariate Gaussian components based on recent developments in weighted likelihood estimation. Actually, the proposed methodology is meant to provide a step further with respect to the original proposal in \cite{markatou2000mixture}. 
The method is based on the idea of using a univariate kernel density estimate based on robust distances rather than a multivariate one based on the data in order to compute weights. %Therefore, we developed a method well suited to handle multivariate data. %, even in large dimensions.
Furthermore, the proposed technique is characterized by %the evaluation of component-wise set of weights and 
the introduction of an eigen constraint aimed at avoiding problems connected with an unbounded likelihood or spurious solutions.  

Based on the robustly fitted mixture model, a model based clustering strategy can be built in a standard fashion by looking at the value of posterior membership probabilities.  
At the same time, formal rules for outlier detection can be derived, as well.
Then, one could assign units to clusters provided that the corresponding outlyingness test is not significant, that means that detected outliers have to be discarded and not assigned to any group.
The numerical studies and the real data examples showed the satisfactory reliability of the proposed methodology. 

There is still room for further work, along a path shared with {\tt tclust}, {\tt rtclust} and {\tt otrimle}. Actually the proposed method works for a given smoothing parameter $h$ and a fixed number of clusters $K$. In addition, outlier detection depends upon a fixed threshold.
At the moment, the selection of $h$ stemming from the monitoring of several quantities, such as the empirical downweighting level, the unit specific robust distances or even the fitted parameters, provides an acceptable adaptive solution. Such a procedure is not different form the
implementation of a sequence of refinement steps of an initial robust partition stemming from a sequence of decreasing values of $h$. 
The selection of $K$ remains a difficult problem to deal with too, despite the satisfactory behavior of the proposed criteria, i.e. the weighted BIC and the weighted Classification log-Likelihood. 
Outlier detection is a novel aspect in the framework of robust mixture modeling and model based clustering. 
%The problem of outlier detection consists in testing $n$ null hypotheses that each data point is a realization of a multivariate normal distribution. 
In the specific context, the outlyingness of each unit is tested conditionally on the final cluster assignment. The number of outliers clearly depends on the chosen level $\alpha$ or the selected threshold for the final weights. A fair choice of the level of the test is still an open problem in outlier detection. However,  the suggested testing strategies work satisfactory, at least in those considered scenarios, and provide a
good compromise between swamping and masking that could be improved further by using multiplicity adjustments \citep{cerioli2011error}.

\begin{table*}
\caption{Average measures of fitting accuracy for WEM, WCEM, EM, CEM and {\tt otrimle} with p=2,5,10, $\epsilon=0$, $\beta=10,8,6$.}
\label{tab1no}
\centering
\begin{tabular}{r|c|ccc|ccc|ccc}
&                &\multicolumn{3}{c|}{$\beta=10$}&\multicolumn{3}{c}{$\beta=8$}&\multicolumn{3}{c}{$\beta=6$}\\
\hline
&         p     &$\mu$&$\Sigma$&$\pi$&$\mu$&$\Sigma$&$\pi$&$\mu$&$\Sigma$&$\pi$\\
\hline
                  &2    &0.600& 0.186&0.022&0.641&0.203&0.023&0.8550&0.248&0.034\\
 WEM        &5    &0.651&0.802&0.025&0.767&0.822&0.035&0.867&0.886&0.032\\
                 &10  &0.664&1.158&0.022&0.703&1.183&0.024&0.921&1.254&0.036\\
\hline
                  &2  &0.614&0.193&0.030&0.687&0.208&0.029&1.273&0.261&0.056\\
  WCEM     &5  &0.653&0.795&0.033&0.767&0.822&0.035&1.422&0.904&0.064\\
                  &10&0.681&1.156&0.028&0.829&1.194&0.031&1.715&1.305&0.074\\     
\hline
                 &2  &0.551&0.154&0.022&0.597&0.169&0.023&0.765&0.184&0.029\\
EM           &5  &0.622&0.764&0.025&0.651&0.713&0.025&0.816&0.832&0.029\\
                 &10&0.659&1.113&0.022&0.697&1.137&0.023&0.933&1.214&0.033\\
\hline
                 &2  &0.603&0.160&0.022&0.803&0.182&0.030&1.760&0.274&0.056\\
CEM         &5  &0.650&0.770&0.028&0.834&0.790&0.033&1.698&0.874&0.073\\
                 &10&0.687&1.122&0.022&0.875&1.158&0.030&1.915&1.270&0.081\\
\hline

                 &2  &0.594&0.170&0.053&0.622 &0.187&0.047&0.773&0.204&0.050\\
{\tt otrimle}&5  &0.636&0.770&0.028&0.666&0.787&0.031&0.834&0.851&0.039\\
                 &10&0.660&1.117&0.023&0.699&1.141&0.024&0.914&1.215&0.033\\
\hline
\end{tabular}
\end{table*}

\begin{table*}[t]
\caption{Swamping rate for WEM, WCEM and {\tt otrimle} with p=2,5,10, $\epsilon=0$, $\beta=10,8,6$.}
\label{tab2no}
\centering
\begin{tabular}{r|c|ccc}
&         p       &$\beta=10$&$\beta=8$&$\beta=6$\\
\hline

                  &2    &0.024& 0.021&0.017\\
 WEM        &5    &0.018&0.017&0.016\\
 $\alpha=0.01$                &10  &0.018&0.017&0.017\\
\hline
                  &2  &0.024&0.020&0.020\\
  WCEM     &5  &0.017&0.016&0.017\\
 $\alpha=0.01$                   &10&0.017&0.017&0.017\\     
\hline
                 &2  &0.007&0.005&0.005\\
WEM           &5  &0.004&0.004&0.004\\
$\hat w<0.1$                 &10&0.004&0.004&0.003\\
\hline
                 &2  &0.007&0.005&0.005\\
WCEM         &5  &0.004&0.003&0.004\\
$\hat w<0.1$                 &10&0.003&0.003&0.003\\
\hline
                 &2  &0.016&0.012&0.010\\
WEM           &5  &0.010&0.009&0.008\\
$\hat w<0.2$                 &10&0.009&0.008&0.008\\
\hline
                 &2  &0.016&0.011&0.011\\
WCEM         &5  &0.008&0.008&0.008\\
$\hat w<0.2$                 &10&0.007&0.007&0.008\\
\hline

                 &2  &0.038&0.028&0.031\\
{\tt otrimle}&5  &0.005&0.007&0.014\\
                 &10&0.019&0.002&0.002\\
\hline
\end{tabular}

\end{table*}

\begin{table*}[t]
\caption{Average measures of classification accuracy for WEM, WCEM, EM, CEM and {\tt otrimle} with p=2,5,10, $\epsilon=0$, $\beta=10,8,6$.}
\label{tab3no}
\centering
\begin{tabular}{r|c|cc|cc|cc}
&                &\multicolumn{2}{c|}{$\beta=10$}&\multicolumn{2}{c}{$\beta=8$}&\multicolumn{2}{c}{$\beta=6$}\\
\hline
&         p     &Rand&MCE&Rand&MCE&Rand&MCE\\
\hline
                  &2    &0.978& 0.007&0.929&0.024&0.829&0.062\\
 WEM        &5    &0.975&0.008&0.929&0.024&0.831&0.062\\
                 &10  &0.974&0.008&0.924&0.026&0.819&0.067\\
\hline
                  &2    &0.979& 0.007&0.933&0.023&0.834&0.061\\
 WCEM        &5    &0.976&0.008&0.930&0.024&0.827&0.064\\
                 &10  &0.974&0.008&0.925&0.026&0.803&0.075\\
\hline
                  &2    &0.973& 0.008&0.928&0.024&0.834&0.060\\
 EM        &5    &0.974&0.008&0.928&0.024&0.832&0.062\\
                 &10  &0.972&0.009&0.923&0.026&0.818&0.067\\
\hline
                  &2    &0.973& 0.009&0.927&0.024&0.817&0.068\\
 CEM        &5    &0.973&0.009&0.927&0.024&0.817&0.068\\
                 &10  &0.971&0.009&0.920&0.027&0.794&0.079\\
\hline

                  &2    &0.975& 0.008&0.930&0.024&0.837&0.059\\
{\tt otrimle} &5    &0.974&0.008&0.930&0.024&0.834&0.061\\
                 &10  &0.973&0.009&0.923&0.026&0.820&0.067\\
\hline
\end{tabular}

\end{table*}

\begin{table*}[t]
\caption{Average measures of fitting accuracy for WEM, WCEM, {\tt tclust} and {\tt otrimle} with p=2,5,10, $\epsilon=0.10,0.20$, $\beta=10$ (well separated clusters).}
\label{tab1}
\centering
\begin{tabular}{r|c|ccc|ccc}
&                &\multicolumn{3}{c|}{$\epsilon=0.10$}&\multicolumn{3}{c}{$\epsilon=0.20$}\\
\hline
&         p     &$\mu$&$\Sigma$&$\pi$&$\mu$&$\Sigma$&$\pi$\\
\hline
                  &2    &0.651& 0.219&0.029&0.745&0.249&0.026\\
 WEM        &5    &0.679&0.886&0.025&0.694&0.894&0.026\\
                 &10  &0.685&1.237&0.023&0.741&1.278&0.027\\
\hline
                  &2  &0.629&0.213&0.035&0.761&0.245&0.059\\
  WCEM     &5  &0.731&0.944&0.045&0.719&1.128&0.045\\
                  &10&0.732&1.301&0.035&0.780&1.642&0.040\\     
\hline
                 &2  &0.619&0.219&0.027&0.774&0.261&0.029\\
{\tt tclust}  &5  &0.631&0.796&0.025&0.685&0.834&0.026\\
                 &10&0.682&1.153&0.023&0.718&1.218&0.028\\
\hline
                 &2  &0.638&0.214&0.110&0.744 &0.239&0.177\\
{\tt otrimle}&5  &0.629&0.803&0.068&0.664&0.845&0.124\\
                 &10&0.668&1.163&0.067&0.704&1.233&0.124\\
\hline
\end{tabular}

\end{table*}

\begin{table*}[t]
\caption{Average measures of fitting accuracy for WEM, WCEM, {\tt tclust} and {\tt otrimle} with p=2,5,10, $\epsilon=0.10,0.20$, $\beta=8$ (moderate overlapping).}
\label{tab2}
\centering
\begin{tabular}{r|c|ccc|ccc}
&                &\multicolumn{3}{c|}{$\epsilon=0.10$}&\multicolumn{3}{c}{$\epsilon=0.20$}\\
\hline
&         p     &$\mu$&$\Sigma$&$\pi$&$\mu$&$\Sigma$&$\pi$\\
\hline
                  &2    &0.684& 0.254&0.029&0.821&0.280&0.030\\
 WEM        &5    &0.727&0.883&0.027&0.742&0.900&0.028\\
                 &10  &0.750&1.233&0.028&0.801&1.308&0.030\\
\hline
                  &2  &0.685&0.230&0.041&1.087&0.285&0.081\\
  WCEM     &5  &0.806&1.170&0.045&0.826&1.128&0.039\\
                  &10&0.858&1.616&0.032&0.928&1.774&0.038\\     
\hline
                 &2  &0.802&0.264&0.034&0.884&0.336&0.039\\
{\tt tclust}  &5  &0.829&0.818&0.033&0.876&0.871&0.033\\
                 &10&0.860&1.187&0.033&0.909&1.258&0.036\\
\hline
                 &2  &0.652&0.238&0.111&0.731&0.276&0.183\\
{\tt otrimle}&5  &0.684&0.812&0.068&0.719&0.872&0.126\\
                 &10&0.731&1.192&0.068&0.782&1.261&0.126\\
\hline
\end{tabular}

\end{table*}

\begin{table*}[t]
\caption{Average measures of fitting accuracy for WEM, WCEM, {\tt tclust} and {\tt otrimle} with p=2,5,10, $\epsilon=0.10,0.20$, $\beta=6$ (severe overlapping).}
\label{tab3}
\centering
\begin{tabular}{r|c|ccc|ccc}
&                &\multicolumn{3}{c|}{$\epsilon=0.10$}&\multicolumn{3}{c}{$\epsilon=0.20$}\\
\hline
&         p     &$\mu$&$\Sigma$&$\pi$&$\mu$&$\Sigma$&$\pi$\\
\hline
                  &2    &1.043& 0.313&0.038&1.131&0.370&0.056\\
 WEM        &5    &0.837&0.946&0.032&0.899&0.977&0.035\\
                 &10  &0.990&1.294&0.040&1.115&1.375&0.046\\
\hline
                  &2  &1.112&0.344&0.058&1.386&0.362&0.081\\
  WCEM     &5  &1.121&1.356&0.072&1.465&1.236&0.054\\
                  &10&1.909&1.756&0.086&1.889&1.951&0.076\\     
\hline
                 &2  &1.728&0.372&0.082&3.431&0.607&0.133\\
{\tt tclust}  &5  &1.477&0.897&0.069&1.685&0.957&0.072\\
                 &10&1.761&1.301&0.077&1.898&1.348&0.083\\
\hline
                 &2  &0.910&0.293&0.122& 0.885&0.295&0.183\\
{\tt otrimle}&5  &0.781&0.876&0.072&0.861&0.923&0.127\\
                 &10&0.991&1.259&0.077&1.086&1.320&0.132\\
\hline
\end{tabular}

\end{table*}

%%%%%%%%%%%%%%%%%%%%%%%%%%%%%%%%%%%%%%%%%
\begin{table*}[t]
\caption{Outlier detection for WEM, WCEM, {\tt tclust} and {\tt otrimle} with p=2,5,10, $\epsilon=0.10,0.20$, $\beta=10$ (well separated clusters). }%The level of the testing procedure stemming from WEM and WCEM is $\alpha=0.01$.}
\label{tab7}
\centering
\begin{tabular}{r|c|ccc|ccc}
&                &\multicolumn{3}{c|}{$\epsilon=0.10$}&\multicolumn{3}{c}{$\epsilon=0.20$}\\
\hline
&         p     &$\epsilon$&swamp.&mask.&$\epsilon$&swamp.&mask.\\
\hline
                  &2    &0.118& 0.021&0.004&0.227&0.034&0.002\\
 WEM        &5    &0.120&0.022&0.000&0.217&0.021&0.000\\
$\alpha=0.01$               &10  &0.119&0.021&0.000&0.218&0.022&0.000\\
\hline
                  &2  &0.118&0.021&0.005&0.223&0.032&0.015\\
  WCEM     &5  &0.118&0.020&0.000&0.217&0.017&0.000\\
$\alpha=0.01$                  &10&0.115&0.017&0.000&0.213&0.016&0.000\\     
\hline
                &2    &0.096& 0.005&0.086&0.211&0.017&0.012\\
 WEM        &5    &0.108&0.009&0.000&0.205&0.006&0.000\\
$w<0.1$               &10  &0.106&0.007&0.000&0.207&0.009&0.000\\
\hline
                  &2  &0.092&0.005&0.125&0.201&0.019&0.069\\
  WCEM     &5  &0.107&0.008&0.000&0.205&0.006&0.001\\
$\hat w<0.1$                  &10&0.105&0.006&0.000&0.206&0.007&0.000\\ 

\hline
                &2    &0.110& 0.013&0.016&0.231&0.039&0.001\\
 WEM        &5    &0.118&0.020&0.000&0.213&0.016&0.000\\
$\hat w<0.2$               &10  &0.116&0.018&0.000&0.216&0.020&0.000\\
\hline
                  &2  &0.108&0.013&0.038&0.227&0.039&0.022\\
  WCEM     &5  &0.116&0.018&0.000&0.211&0.014&0.000\\
$\hat w<0.2$                  &10&0.114&0.016&0.001&0.213&0.016&0.000\\
    
\hline
                 &2    &0.109& 0.012&0.018&0.263&0.078&0.000\\
 WEM        &5    &0.126&0.031&0.000&0.224&0.031&0.000\\
$\hat w<1-\hat{\bar w}$               &10  &0.122&0.024&0.000&0.238&0.048&0.000\\
\hline
                  &2  &0.109&0.013&0.016&0.263&0.080&0.007\\
  WCEM     &5  &0.125&0.028&0.000&0.224&0.030&0.000\\
$\hat w<1-\hat{\bar w}$                  &10&0.120&0.025&0.000&0.232&0.040&0.000\\     
\hline

                 &2  &0.100&0.005&0.043&0.200&0.008&0.031\\
{\tt tclust}  &5  &0.100&0.000&0.000&0.200&0.000&0.000\\
                 &10&0.100&0.000&0.000&0.200&0.000&0.000\\
\hline
                 &2  &0.135&0.040&0.019&0.254 &0.069&0.005\\
{\tt otrimle}&5  &0.106&0.003&0.007&0.203&0.003&0.004\\
                 &10&0.102&0.002&0.001&0.202&0.003&0.000\\
\hline
\end{tabular}

\end{table*}

\begin{table*}[t]
\caption{Outlier detection for WEM, WCEM, {\tt tclust} and {\tt otrimle} with p=2,5,10, $\epsilon=0.10,0.20$, $\beta=8$ (moderate overlapping).} %The level of the testing procedure stemming from WEM and WCEM is $\alpha=0.01$.}
\label{tab8}
\centering
\begin{tabular}{r|c|ccc|ccc}
&                &\multicolumn{3}{c|}{$\epsilon=0.10$}&\multicolumn{3}{c}{$\epsilon=0.20$}\\
\hline
&         p     &$\epsilon$&swamp.&mask.&$\epsilon$&swamp.&mask.\\
\hline
                  &2    &0.120& 0.023&0.001&0.218&0.025&0.012\\
 WEM        &5    &0.126&0.029&0.000&0.216&0.021&0.000\\
 $\alpha=0.01$                 &10  &0.125&0.028&0.000&0.217&0.021&0.000\\
\hline
                  &2  &0.117&0.024&0.048&0.216&0.029&0.038\\
  WCEM     &5  &0.127&0.031&0.000&0.214&0.017&0.000\\
  $\alpha=0.01$                 &10&0.123&0.026&0.000&0.213&0.016&0.000\\     
\hline
       &2    &0.096& 0.005&0.086&0.202&0.012&0.038\\
 WEM        &5    &0.108&0.009&0.000&0.205&0.006&0.000\\
$\hat w<0.1$               &10  &0.106&0.007&0.000&0.207&0.009&0.000\\
\hline
                  &2  &0.092&0.005&0.125&0.198&0.019&0.085\\
  WCEM     &5  &0.107&0.008&0.000&0.205&0.006&0.001\\
$\hat w<0.1$                  &10&0.105&0.006&0.000&0.205&0.016&0.000\\    

\hline
                &2    &0.110& 0.013&0.016&0.221&0.028&0.009\\
 WEM        &5    &0.118&0.020&0.000&0.212&0.014&0.000\\
$\hat w<0.2$               &10  &0.116&0.018&0.000&0.215&0.019&0.000\\
\hline
                  &2  &0.108&0.013&0.038&0.225&0.041&0.042\\
  WCEM     &5  &0.116&0.018&0.000&0.211&0.013&0.000\\
$\hat w<0.2$                  &10&0.114&0.016&0.001&0.212&0.016&0.000\\
 
\hline
                 &2    &0.109& 0.012&0.018&0.250&0.063&0.002\\
 WEM        &5    &0.126&0.031&0.000&0.223&0.029&0.000\\
$\hat w<1-\hat{\bar w}$               &10  &0.122&0.024&0.000&0.236&0.046&0.000\\
\hline
                  &2  &0.109&0.013&0.016&0.272&0.097&0.025\\
  WCEM     &5  &0.125&0.028&0.000&0.222&0.028&0.000\\
$\hat w<1-\hat{\bar w}$                  &10&0.120&0.025&0.000&0.231&0.039&0.000\\     
\hline
                 &2  &0.100&0.007&0.064&0.200&0.009&0.036\\
{\tt tclust}  &5  &0.100&0.000&0.001&0.200&0.000&0.001\\
                 &10&0.100&0.000&0.000&0.200&0.000&0.000\\
\hline
                 &2  &0.141&0.047&0.005&0.256 &0.069&0.000\\
{\tt otrimle}&5  &0.101&0.001&0.007&0.206&0.008&0.004\\
                 &10&0.102&0.002&0.001&0.203&0.004&0.000\\
\hline
\end{tabular}

\end{table*}

\begin{table*}[t]
\caption{Outlier detection for WEM, WCEM, {\tt tclust} and {\tt otrimle} with p=2,5,10, $\epsilon=0.10,0.20$, $\beta=6$ (severe overlapping).} %The level of the testing procedure stemming from WEM and WCEM is $\alpha=0.01$.}
\label{tab9}
\centering
\begin{tabular}{r|c|ccc|ccc}
&                &\multicolumn{3}{c|}{$\epsilon=0.10$}&\multicolumn{3}{c}{$\epsilon=0.20$}\\
\hline
&         p     &$\epsilon$&swamp.&mask.&$\epsilon$&swamp.&mask.\\
\hline
                  &2    &0.120& 0.023&0.001&0.206&0.020&0.052\\
 WEM        &5    &0.125&0.028&0.000&0.214&0.018&0.000\\
 $\alpha=0.01$                 &10  &0.123&0.026&0.000&0.216&0.020&0.000\\
\hline
                  &2  &0.145&0.050&0.001&0.253&0.045&0.006\\
  WCEM     &5  &0.128&0.031&0.000&0.214&0.018&0.000\\
  $\alpha=0.01$                 &10&0.124&0.027&0.000&0.213&0.017&0.000\\     
\hline
       &2    &0.096& 0.005&0.086&0.209&0.020&0.003\\
 WEM        &5    &0.108&0.009&0.000&0.205&0.006&0.000\\
$\hat w<0.1$               &10  &0.106&0.007&0.000&0.207&0.008&0.000\\
\hline
                  &2  &0.092&0.005&0.125&0.237&0.046&0.001\\
  WCEM     &5  &0.107&0.008&0.000&0.206&0.006&0.001\\
$\hat w<0.1$                  &10&0.105&0.006&0.000&0.206&0.007&0.000\\    

\hline
                &2    &0.110& 0.013&0.016&0.231&0.039&0.001\\
 WEM        &5    &0.118&0.020&0.000&0.211&0.013&0.000\\
$\hat w<0.2$               &10  &0.116&0.018&0.000&0.215&0.019&0.000\\
\hline
                  &2  &0.108&0.013&0.038&0.227&0.039&0.022\\
  WCEM     &5  &0.116&0.018&0.000&0.211&0.013&0.001\\
$\hat w<0.2$                  &10&0.114&0.016&0.001&0.213&0.016&0.000\\
 
\hline
                 &2    &0.109& 0.012&0.018&0.209&0.020&0.003\\
 WEM        &5    &0.126&0.031&0.000&0.222&0.027&0.000\\
$\hat w<1-\hat{\bar w}$               &10  &0.122&0.024&0.000&0.235&0.044&0.000\\
\hline
                  &2  &0.109&0.013&0.016&0.237&0.046&0.001\\
  WCEM     &5  &0.125&0.028&0.000&0.223&0.028&0.000\\
$\hat w<1-\hat{\bar w}$                  &10&0.120&0.025&0.000&0.232&0.041&0.000\\     
\hline
                 &2  &0.100&0.007&0.061&0.200&0.024&0.095\\
{\tt tclust}  &5  &0.100&0.000&0.002&0.200&0.000&0.001\\
                 &10&0.100&0.000&0.000&0.200&0.000&0.000\\
\hline
                 &2  &0.145&0.050&0.004&0.251 &0.064&0.003\\
{\tt otrimle}&5  &0.102&0.003&0.006&0.202&0.003&0.006\\
                 &10&0.102&0.003&0.001&0.203&0.004&0.000\\
\hline
\end{tabular}

\end{table*}

%%%%%%%%%%%%%%%%%%%%%%%%%%%%%%%%%%%%%%%%%%%

\begin{table*}[t]
\caption{Average measures of classification accuracy for WEM, WCEM, {\tt tclust} and {\tt otrimle} with p=2,5,10, $\epsilon=0.10,0.20$, $\beta=10$  (well separated clusters).}
\label{tab4}
\centering
\begin{tabular}{r|c|cc|cc}
&                &\multicolumn{2}{c|}{$\epsilon=0.10$}&\multicolumn{2}{c}{$\epsilon=0.20$}\\
\hline
&         p     &Rand&MCE&Rand&MCE\\
\hline
                  &2    &0.98& 0.01&0.98&0.01\\
 WEM        &5    &0.97& 0.01&0.97&0.01\\
                 &10  &0.97& 0.01&0.97&0.01\\
\hline
                  &2  &0.98& 0.01&0.97&0.01\\
  WCEM     &5  &0.97& 0.01&0.97&0.01\\
                  &10&0.97& 0.01&0.97&0.01\\    
\hline
                 &2  &0.97& 0.01&0.97&0.01\\
{\tt tclust}  &5 &0.97& 0.01&0.97&0.01\\
                 &10&0.97& 0.01&0.97&0.01\\
\hline
                 &2  &0.98& 0.01&0.98&0.01\\
{\tt otrimle}&5  &0.97& 0.01&0.97&0.01\\
                 &10&0.97& 0.01&0.97&0.01\\
\hline
\end{tabular}

\end{table*}

\begin{table*}[t]
\caption{Average measures of classification accuracy for WEM, WCEM, {\tt tclust} and {\tt otrimle} with p=2,5,10, $\epsilon=0.10,0.20$, $\beta=8$  (moderate overlapping).}
\label{tab5}
\centering
\begin{tabular}{r|c|cc|cc}
&                &\multicolumn{2}{c|}{$\epsilon=0.10$}&\multicolumn{2}{c}{$\epsilon=0.20$}\\
\hline
&         p     &Rand&MCE&Rand&MCE\\
\hline
                  &2    &0.93& 0.02&0.92&0.03\\
 WEM        &5    &0.92& 0.03&0.93&0.03\\
                 &10  &0.92& 0.03&0.92&0.03\\
\hline
                  &2  &0.93& 0.02&0.91&0.03\\
  WCEM     &5  &0.92& 0.03&0.93&0.03\\
                  &10&0.92& 0.03&0.92&0.03\\    
\hline
                 &2  &0.93& 0.02&0.92&0.03\\
{\tt tclust}  &5 &0.92& 0.03&0.93&0.03\\
                 &10&0.92& 0.03&0.92&0.03\\
\hline
                 &2  &0.93& 0.02&0.93&0.02\\
{\tt otrimle}&5  &0.92& 0.03&0.93&0.03\\
                 &10&0.92& 0.03&0.92&0.03\\
\hline
\end{tabular}

\end{table*}

\begin{table*}[t]
\caption{Average measures of classification accuracy for WEM, WCEM, {\tt tclust} and {\tt otrimle} with p=2,5,10, $\epsilon=0.10,0.20$, $\beta=6$  (severe overlapping).}
\label{tab6}
\centering
\begin{tabular}{r|c|cc|cc}
&                &\multicolumn{2}{c|}{$\epsilon=0.10$}&\multicolumn{2}{c}{$\epsilon=0.20$}\\
\hline
&         p     &Rand&MCE&Rand&MCE\\
\hline
                  &2    &0.82& 0.07&0.79&0.07\\
 WEM        &5    &0.82& 0.06&0.82&0.06\\
                 &10  &0.82& 0.07&0.81&0.07\\
\hline
                  &2  &0.82& 0.06&0.79&0.08\\
  WCEM     &5  &0.81& 0.07&0.81&0.07\\
                  &10&0.79& 0.08&0.79&0.08\\    
\hline
                 &2  &0.82& 0.07&0.76&0.07\\
{\tt tclust}  &5 &0.82& 0.07&0.82&0.07\\
                 &10&0.80& 0.08&0.80&0.08\\
\hline
                 &2  &0.84& 0.06&0.83&0.06\\
{\tt otrimle}&5  &0.83& 0.06&0.83&0.06\\
                 &10&0.82& 0.07&0.81&0.07\\
\hline
\end{tabular}

\end{table*}
%%%%%%%%%%%%%%%%%%%%%%%%%%%%%%%%%%%%%%%%%%%

\begin{table*}[t]
\label{tab_whr}
\caption{WHR18 data: cluster profiles and raw measurements for the detected outlying countries.}
\centering
\scriptsize
\begin{tabular}{rccccccc}
  \hline
 & LogGDP & HLE & Social support & Freedom & Generosity & Corruption & Size \\ 
  \hline
  Cluster 1 & 7.88 & 53.68 & 0.69 & 0.69 & 0.19 & 0.78 &38\\
  Cluster 2 & 9.35 & 64.46 & 0.83 & 0.74 & 0.26 & 0.80 &60\\ 
  Cluster 3 & 10.48 & 71.63 & 0.90 & 0.83 & 0.44 & 0.61&37 \\   
  \hline
  Central African Rep. & 6.47 & 44.31 & 0.31 & 0.63 & 0.17 & 0.88& \\ 
  Chad & 7.55 & 45.66 & 0.68 & 0.53 & 0.17 & 0.84 &\\ 
    Ivory Coast & 8.13 & 46.52 & 0.66 & 0.77 & 0.16 & 0.76 &\\ 
  Lesotho & 7.91 & 46.48 & 0.80 & 0.73 & 0.10 & 0.74 &\\ 
  Myanmar & 8.59 & 57.51 & 0.79 & 0.86 & 0.90 & 0.62& \\ 
  Nigeria & 8.61 & 45.50 & 0.78 & 0.76 & 0.28 & 0.89 &\\ 
  Sierra Leone & 7.22 & 43.99 & 0.64 & 0.67 & 0.24 & 0.85& \\ 
   \hline
\end{tabular}
\end{table*}

\bibliographystyle{spbasic}
\bibliography{biblioEM}
\end{document}